\documentclass{emulateapj}
\def\stacksymbols #1#2#3#4{\def\theguybelow{#2}
        \def\verticalposition{\lower#3pt}
        \def\spacingwithinsymbol{\baselineskip0pt\lineskip#4pt}
        \mathrel{\mathpalette\intermediary#1}}
\def\intermediary #1#2{\verticalposition\vbox{\spacingwithinsymbol
        \everycr={}\tabskip0pt
        \halign{$\mathsurround0pt#1\hfil##\hfil$\crcr#2\crcr
                \theguybelow\crcr}}}
\def\lta{\stacksymbols{<}{\sim}{2.5}{.2}}
\def\gta{\stacksymbols{>}{\sim}{3}{.5}}

\shorttitle{FIR Emission from Elliptical Galaxies}
\shortauthors{Temi, Brighenti \& Mathews}

\begin{document}

\title{FAR INFRARED {\it SPITZER} OBSERVATIONS OF ELLIPTICAL GALAXIES: 
EVIDENCE FOR EXTENDED DIFFUSE DUST}

\author{Pasquale Temi\altaffilmark{1,2}, 
Fabrizio Brighenti\altaffilmark{3,4}, William G. Mathews\altaffilmark{3} }
%\email{ptemi@mail.arc.nasa.gov}
%\email{mathews@ucolick.org}
%\email{fabrizio.brighenti@unibo.it}

\altaffiltext{1}{Astrophysics Branch, NASA/Ames Research Center, MS 245-6,
Moffett Field, CA 94035.}
\altaffiltext{2}{SETI Institute, Mountain View, CA 94043;
and Department of Physics and Astronomy, University of Western Ontario, 
London, ON N6A 3K7, Canada.}
\altaffiltext{3}{University of California Observatories/Lick Observatory,
Board of Studies in Astronomy and Astrophysics,
University of California, Santa Cruz, CA 95064.}
\altaffiltext{4}{Dipartimento di Astronomia,
Universit\`a di Bologna, via Ranzani 1, Bologna 40127, Italy.}

\begin{abstract}
Far-infrared {\it Spitzer} observations of elliptical galaxies
are inconsistent with simple steady state models of dust
creation in red giant stars and destruction by grain sputtering
in the hot interstellar gas at $T \sim 10^7$ K.
The flux at 24$\mu$m correlates with optical fluxes, 
suggesting that this relatively hot dust is largely
circumstellar. But fluxes at 70$\mu$m and 160$\mu$m 
do not correlate with optical fluxes. 
Elliptical galaxies with similar $L_B$ have 
luminosities at 70$\mu$m and 160$\mu$m ($L_{70}$ and $L_{160}$) 
that vary over a factor $\sim 100$, implying an additional source
of dust unrelated to that produced by ongoing local stellar mass loss.
Neither $L_{70}/L_B$ nor $L_{160}/L_B$ correlate with
the stellar age or metallicity.
Optical line fluxes from warm gas at $T \sim 10^4$ K 
correlate weakly with $L_{70}$ and $L_{160}$, suggesting that
the dust may be responsible for cooling this gas.
Many normal elliptical galaxies have emission at 70$\mu$m that is
extended to 5--10 kpc.
Extended far-infrared emission with sputtering lifetimes of
$\sim 10^8$ yrs is difficult to maintain by mergers
with gas-rich galaxies.
Instead, we propose that this cold dust is buoyantly transported
from reservoirs of dust in the galactic cores which is 
supplied by mass loss from stars in the core. 
Intermittent energy outbursts from AGNs can drive the buoyant 
outflow.
\end{abstract}


\section{Introduction}

The study of far-infrared (FIR) emission from elliptical galaxies 
is particularly interesting because 
the internal rates that dust is formed and destroyed 
can both be estimated. 
Dust is produced by mass loss from evolving red giant stars in 
the old stellar population typical of these galaxies. 
This mass loss is directly observed in the cumulative mid-infrared 
10--24$\mu$m 
emission from circumstellar regions around giant stars, 
causing the surface brightness profiles at 10--24$\mu$m 
to resemble those at optical wavelengths 
(Knapp et al. 1989; Athey et al. 2002).
The dust destruction rate by sputtering can be estimated from the 
frequency of thermal impacts from ions in the hot interstellar gas 
which for many ellipticals can be estimated from X-ray observations. 
During their sputtering lifetimes, dust grains ejected from giant 
stars are heated by starlight and thermal electron impacts to 
temperatures that radiate in the FIR 
(e.g. Tsai \& Mathews 1995, 1996; Temi et al. 2003). 
This radiation has been observed both by the {\it Infrared 
Astronomical Satellite, IRAS} (Knapp et al. 1989) and by 
the {\it Infrared Space Observatory, ISO} (Temi et al. 2004).
In addition, 
small optically obscuring clouds and disks of dusty gas have 
been detected in about half of the bright ellipticals studied 
with the {\it Hubble Space Telescope, HST}
(van Dokkum \& Franx 1995; Ferrari et al. 1999; Lauer et al. 2005).
These clouds probably formed from dusty gas ejected from 
stars within the central $\sim$kpc in these elliptical galaxies 
(Mathews \& Brighenti 2003).

Galaxy mergers are 
another obvious possible source of dust in elliptical galaxies, 
indeed it is widely thought that ellipticals are formed 
by mergers. 
Elliptical-looking galaxies can be produced after a few Gyrs 
following the merger of two disk galaxies (e.g. Barnes \& Hernquist
1992), but this simple recipe may not easily explain all structural 
and kinematic properties of these galaxies 
(Naab, Jesseit \& Burkert 2006). 
Many of the mergers that produced the ripples, shells and
counter-rotating cores in elliptical galaxies probably occurred 
a very long time ago. 
It is likely that massive boxy ellipticals, similar to many of 
those with strong FIR emission, have 
experienced mostly ``dry'' gas and dust-free mergers during the last few 
Gyrs (Bell et al. 2004; Faber et al. 2006).
Nevertheless, mergers do occasionally occur between 
low-redshift ellipticals and gas-rich 
galaxies and 
these post-merger ellipticals often selectively appear in 
the observing lists of infrared observers. 
Our previous study of FIR emission from elliptical galaxies 
was based on a sample chosen from 
the {\it ISO} data archive (Temi et al. 2004) that included a
significant number of galaxies with abnormally large masses of 
dust and cold gas which are uncharacteristic of elliptical galaxies 
in general.  
The large range of FIR luminosities that we found 
in the {\it ISO} sample was therefore not unexpected, and 
it was reasonable to suppose that mergers with dusty 
galaxies explained the 
absence of correlations between optical and FIR emission.
The failure to find correlations between the FIR and optical 
luminosities of elliptical galaxies has been used in the past to 
argue that the cold dust is acquired in a stochastic fashion by
mergers (Forbes 1991).

To understand the relative role of internal and external sources 
of dust in elliptical galaxies, we describe here 
new {\it Spitzer} observations at 24, 70 and 160$\mu$m.
Our full sample consists of 46 elliptical galaxies, formed 
by combining our own proprietary {\it Spitzer} 
observations with those of other observers. 
The improved sensitivity 
and angular resolution of {\it Spitzer} provides clear 
evidence that the 70$\mu$m emission is spatially 
extended in many ellipticals.  
In particular, we find extended FIR emission in elliptical galaxies
that do not contain uncharacteristically 
large masses of neutral, molecular or 
optically visible cold gas. 
The diffuse nature of the dust in these otherwise normal galaxies, 
combined with the relatively short (ion sputtering) lifetime for 
dust destruction in 
the hot gas, favors an internal source of dust 
rather than dust-rich mergers which are probably too infrequent. 
The luminosity of extended FIR dust 
emission in most ellipticals greatly exceeds 
that expected from normal mass loss from local stars. 
However, small, 
optically thick nuclear dust clouds observed in many elliptical
galaxies may provide an alternative, more plausible source for the 
extended FIR dust emission. 
Consequently, we argue that dust accumulated in the central regions of 
elliptical galaxies from stellar sources in the core is 
transported out to larger radii by 
buoyantly rising gas intermittently heated by low-luminosity active 
galactic nuclei (AGN) in the galactic cores. 
If this hypothesis can be confirmed in more detail,
the presence of extended cold dust can be used to trace the 
radial circulation of hot gas in these galaxy/groups. 
Such circulation flows are known to be consistent with X-ray 
observations in which very little hot gas cools to low temperatures 
(Mathews, Brighenti \& Buote 2004).

\section{The Sample}

For this study we construct a sample of 46 elliptical galaxies 
by combining far-infrared {\it Spitzer} observations from our own 
{\it Spitzer} data with additional galaxies in the public 
{\it Spitzer} archive.
Our proprietary far-infrared {\it Spitzer} 
observations of elliptical galaxies (PID 20171, PI P.Temi) 
consists of 11 non-peculiar, bright 
($L_B\geq 2\times 10^{10}$ $L_{B,\odot}$) elliptical galaxies, 
most having high-quality X-ray observations. 
Galactic proximity and spatial scale were also important factors in 
choosing this sample since high angular resolution is needed to separate 
emission from dusty cores and diffuse emission from truly 
interstellar dust spread over the galactic volume.
Furthermore, we avoided sources with strong radio sources and AGN 
that have detectable signals in the far infrared. 
Ellipticals 
with strong infrared Milky Way cirrus or bright spiral galaxies in the 
vicinity were also avoided. 
Since one of our goals is to 
study the interaction between dust and hot gas, we included many 
galaxies for which the X-ray emission is well-studied by either 
{\it Chandra} or {\it XMM}. 

In addition to our original sample we selected 35 
additional ellipticals 
from data currently available in the public {\it Spitzer} archive 
targeted by several PIs during their General
Observation in Cycles 1 and 2, Guaranteed Time Observations, 
and the Legacy Program SINGS (Kennicutt et al. 2003). 
All the galaxies added to our original sample were 
chosen using selection criteria tailored to our specific 
scientific goals. Thus, the 46 galaxies studied here form a rather
homogeneous sample and is large enough to be statistically meaningful.

Tables 1 and 2 summarize the relevant properties of the sample
galaxies including the far-infrared photometric measurements. 
Optical data are taken from the LEDA catalog (Paturel et al. 1997)
and the RSA catalog (Sandage \& Tamman 1981), while X-ray data are
from Ellis and O'Sullivan et al. (2006). 
Luminosities and distances are calculated
with $H_0=70$ km s$^{-1}$ Mpc$^{-1}$ and are derived from the flux
densities in Table 1 using 
far-infrared 
(FIR) passbands of $\Delta \lambda = $ 5.3, 
19 and 34$\mu$m for emission at 24, 70 and 160$\mu$m respectively.
The PIs of the original Spitzer program from which the galaxies have been 
selected are shown in column two of Table 1. All galaxies are 
classified as ellipticals 
in the NED catalog, 
and the morphological type T is taken from the LEDA catalog. 
Nevertheless, we shall find that our sample contains some S0 galaxies.
Most of the sample galaxies have known X-ray luminosities
$L_x$, which are a measure of the hot gas present in each galaxy.

Of the 46 galaxies in the sample three are 
dwarf elliptical galaxies (NGC855, NGC1377, and NGC3265)
with $L_B \le 5 \times 10^9$ L$_{B,\odot}$. Although they are listed
in Table 1, they are not included in our discussion.

\section{Observations and data reduction}

The data presented here were acquired with the Multiband Imager
Photometer (MIPS)
(Rieke et al. 2004) on the {\it Spitzer Space Telescope} (Werner
et al. 2004) in three wavebands centered at 24, 70 and 
160$\mu m$.
Since the galaxy sample has been constructed by collecting archival data
from several Spitzer observing programs, the image data were recorded
using observing modes and integrations times tailored for their original
scientific goals.
As a consequence, although the maps represent the deepest FIR
images acquired to date, some data achieve sensitivities
that are not always optimized for targets that are intrinsically
faint at far-infrared wavelengths.
For example, three of the 
brightest ellipticals in the Virgo Cluster (NGC4472, NGC4636, and NGC4649)
recorded under a guaranteed time program (PID 69, PI G. Fazio)
reach a relatively low sensitivity of 0.5MJy/sr and 1.1MJy/sr 
(1$\sigma$) at 70 and 160$\mu m$, respectively, while other 
galaxies
(i.e. PID 20171, PI P. Temi) have deeper maps at a sensitivity level of
only 0.12MJy/sr and 0.3MJy/sr for the same two wavebands. 
The SINGS data are recorded in MIPS scan mode, covering
a very large sky area ($\sim 30^\prime \times 10^\prime$), 
incorporating two separated passes at each 
source location. SINGS images correspond to maps with intermediate
sensitivity. Details on 
the observing strategies, field coverage and integration times for
the SINGS program are described by Kennicutt et al. (2003).
Apart from the SINGS observations, data have been acquired in MIPS
photometry mode, allowing appropriate coverage of the sources and 
their extended emission. At all wavelengths,
sufficiently large field sizes were chosen 
to reliably measure the background emission.
We used the  Basic Calibrated Data (BCD) products from the Spitzer
Science pipeline (version 13.2) to construct mosaic
images for all objects. Pipeline reduction and post-BCD processing
using the MOPEX software package (Makovoz et al. 2006) 
provide all necessary steps
to process individual frames: dark subtraction,
flat-fielding, mux-bleed correction, flux calibration, correction
of focal plane geometrical distortion, and cosmic ray rejection.

Foreground stars and background galaxies were present in the final 
mosaiced images at all bands. These were identified by eye
and cross-checked using surveys at other
wavelengths (Digital Sky Survey and 2MASS).
They were then masked out from each MIPS image before flux 
extraction was performed.
Also, a few asteroids were identified and removed from the affected maps.
Aperture photometry was performed by measuring
fluxes in suitable circular apertures around the centers.
Flux densities were extracted from apertures extending out 
to the entire optical disk of each galaxy ($R_{25}$).
Sky subtraction was performed by averaging values from multiple
apertures placed around the target, avoiding any overlap
with the faint extended emission from the galaxy.
Uncertainties in flux densities due to sky subtraction
are generally less than 1\%, 
but can be of the order of tens of percent for faint sources.

The measured flux densities are listed in Table 1. Most
of the galaxies in the sample have been detected by MIPS, and 
among them a large fraction show far-infrared emission well beyond
the sensitivity capabilities of previous infrared space observatories.
Although most of the galaxies presented in this paper are unpublished
either because they are our own proprietary data or because the analysis 
is still in progress by the original PIs, results for a 
few galaxies are already
present in the literature (i.e. the SINGS galaxies). We find that our
flux densities are in good agreement with previously published data.
The only exception is the measured flux density at 160$\mu$m reported 
by Dale et al. (2005) for the SINGS galaxy NGC584. Very likely this 
discrepancy is 
due to a misidentification of the target in their very large 160$\mu$m
scan map.
Corrections for extended emission were applied
to the fluxes as described in the Spitzer Observer's manual.
Photometric errors listed in Table 1 refer only to the
statistical uncertainties. Total errors should include 
systematic uncertainties due to flux calibration and to
sky subtraction errors.
The uncertainties on the final absolute calibration are estimated
at 10\% for the 24$\mu$m data, and 20\% at 70 and 160$\mu m$.

Figure 1 shows gray-scale images of a representative sub-sample of 
galaxies. The column of panels on the left is the 24$\mu$m data,
while 70 and 160$\mu$m data are shown in the center and right columns,
respectively. Galaxies are ordered by ascending IR flux in a common
field of view of $3^{\prime} \times 3^{\prime}$. Some very faint
galaxies are shown, including a non detection at 160$\mu$m
and few sources with extended far-infrared emission.
Detailed images of the diffuse extended emission 
in NGC4636 and NGC5044 will be presented in a companion
paper currently in preparation. 

\section{Comparison of FIR, optical and X-ray Emission}

In Figure 2 we compare fluxes at 24, 70 and 160$\mu$m with 
optical B-band fluxes and a related 
plot of FIR and B-band 
luminosities is shown in Figure 3.
Both the 24$\mu$m and B-band fluxes and luminosities 
are clearly correlated,
\begin{equation}
%\log F_{24} = (8.4 \pm 0.4) + (0.72 \pm 0.04)\log F_{B,\odot}
\log F_{24} = (8.37 \pm 0.42) + (0.716 \pm 0.043)\log F_{B,\odot}
\end{equation}
and 
\begin{equation}
%\log L_{24} = (32.3 \pm 0.7) + (0.79 \pm 0.07)\log L_{B,\odot}.
\log L_{24} = (32.37 \pm 0.69) + (0.799 \pm 0.066)\log L_{B,\odot}.
\end{equation}
The approximate consistency with unit slope $F_{24} \propto F_B$ 
indicates that the emission at 24$\mu$m, like that at mid-IR 
wavelengths (Knapp et al. 1989; Athey et al. 2002), 
is dominated by circumstellar 
emission that closely traces optical starlight. 

However, the huge scatter seen in the 70$\mu$m--Bband 
panels in Figures 2 and 3 
suggest that this colder dust is unassociated with stars and is truly 
interstellar. 
The absence of any clear correlation between FIR and optical emission 
from ellipticals has been discussed by 
Forbes (1991), Goudfrooij \& de Jong (1995), 
Trinchieri \& Goudfrooij (2002) and Temi et al. (2004).
In principle cold, FIR-emitting dust could either be diffusely distributed 
in the hot gas or 
confined to dense central clouds similar to those visible 
in optical absorption (e.g. Lauer et al. 2005).
We discuss below our preference for the diffuse component 
in producing most of the FIR-emission. 

It is remarkable that $L_{70}$ and 
$L_{160}$ range over more than two orders of magnitude for galaxies 
with similar $L_B$ -- this is a significant new result 
that must be explained. 
In our previous survey of 
archival data from {\it Infrared Space Observatory} (ISO) 
we found a similar 
spread in FIR luminosities 
at 60-200$\mu$m, but the FIR luminosities in the ISO data 
spanned only one order of magnitude for fixed $L_B$ 
(Temi et al. 2004). 
At the time we attributed much of this scatter to the unusual 
nature of the ISO sample which contained many (post-merger) elliptical 
galaxies with massive disks of cold, dusty gas. 
Now with this {\it Spitzer} sample 
of (mostly) normal elliptical galaxies 
we find that the range
of FIR luminosities is even larger, although we show in Appendix A 
that some of the FIR-brightest sample galaxies cannot be regarded 
as normal. 
In Figure 4 we show that there is no obvious trend of dust color 
$F_{70}/F_{160}$  
(i.e. mean dust temperature) with $L_{70}$ 
except perhaps a tendency for cooler dust in more FIR-luminous galaxies. 
A similar conclusion results if we replace 
$F_{70}$ on the horizontal axis with $F_{160}$, 
$L_{70}$ or $L_{160}$. 

Figures 5 and 6 show no discernable correlation of either $L_{70}/L_B$ or 
$L_{160}/L_B$ with stellar Balmer line ages or metallicity. 
However, $L_{24}/L_B$ may show a slight tendency to decrease 
with age 
\begin{equation}
\log L_{24}/L_B = (30.35 \pm 0.05) - (0.021 \pm 0.005){\rm age}
\end{equation}
or increase with metallicity
\begin{equation}
\log L_{24}/L_B = (29.95 \pm 0.06) + (0.7 \pm 0.2)\log[Z/H].
\end{equation}
A negative correlation of $L_{24}/L_B$ with age would be expected 
since the rate of mass loss from evolving stars 
decreases with stellar age (Piovan et al. 2003), although 
this expected trend is not supported by our study of 
elliptical galaxies at mid-IR wavelengths (Temi et al. 2005). 
It must be noted however that Balmer line ages for the same galaxy 
often show considerable variation when determined by different
authors, so the anticorrelation 
of $L_{24}/L_B$ with age needs further confirmation. 
A positive correlation of $L_{24}/L_B$ with metallicity 
in Figure 6 is 
expected if more interstellar grains are produced by mass-losing 
stars of higher metallicity. 
However, this is inconsistent with the negative correlation at 
70$\mu$m in Figure 6,
\begin{equation}
\log L_{70}/L_B = (30.7 \pm 0.1) - (1.9 \pm 0.4)\log[Z/H].
\end{equation}

No correlations with X-ray luminosity with  
any of the FIR luminosities are visible in Figure 7.
Decreasing FIR luminosities with increasing hot gas density 
might be expected since interstellar grains are destroyed by
sputtering which is proportional to the density of the hot gas. 
However, the relationship between $L_x$ and gas density is unclear. 
For example, there is no correlation between $L_x$ and the 
hot gas density at $\sim 10$ kpc in the 
elliptical galaxies observed with 
{\it Chandra} by Humphrey et al. (2006). 
This disconnect between gas density and $L_x$ may arise because 
the hot gas density on kpc-galactic scales is influenced 
by intermittent energy supplied by the central black holes, 
while $L_x$ is largely determined by more extensive gas 
that fills the dark halos 
that surround the elliptical galaxies 
(Mathews et al. 2006). 
Consequently, the absence of correlations in Figure 7 is 
not surprising. 
Finally, we note that all qualitative features in Figure 7 are 
preserved if luminosities are replaced with fluxes. 

In Figure 8 we compare FIR fluxes with optical emission 
line fluxes determined by three different observers: 
Goudfrooij et al. (1994a,b), Macchetto et al. (1996) and 
Sarzi et al. (2006). 
The optical observations require difficult background subtraction 
and the procedures used vary among the observers. 
For this reason, we compare FIR fluxes 
in Figure 8 with each set of optical emission line 
fluxes independently. 

We believe that both $F_{60}$ and $F_{170}$ correlate weakly with 
observed optical line fluxes, but that $F_{24}$ does not. 
The recent SAURON data for $F_{H\beta}$ 
(Sarzi et al. 2006) show the 
tightest relationship. 
The Goudfrooij et al. data (left column of panels in Figure 8) 
are marginally correlated, but the 
significance increases if it is combined with the Macchetto et al. 
data (shown in the central column).

In Figure 9 we compare the fluxes in the optical emission
lines with B-band fluxes, using only data for non-peculiar
E or E/S0 galaxies, excluding peculiar and known S0 galaxies.
The top and bottom panels from Macchetto et al. (1996) and
Sarzi et al. (2006) show that the line flux is weakly
correlated with the stellar continuum
(see also Phillips et al. 1986; Macchetto et al. 1996;
Goudfrooij 1999).
A correlation between extended line emission and the optical
continuum suggests that the warm gas with $T \sim 104$ K
has a stellar origin.
But the quality of these difficult optical observations is
poor. There are very few objects in common among the
Goudfrooij, Macchetto and Sarzi data sets,
but the flux-flux plots of these common galaxies show
large scatter with almost no clear monotonic relationship.
Five of the six galaxies in common between Goudfrooij and
Macchetto correlate {\it negatively} in a flux-flux plot. 

The correlation in Figure 9, if real, 
may simply indicate that the warm gas and the
dust associated with it both have a stellar origin -- more stars
of a similar age and metallicity eject more gas.
However, the optical emission line flux
depends sensitively on the
density of the warm gas and therefore on the 
pressure in the local hot interstellar gas.
At present there are too few accurate observations to interpret
Figure 9 with confidence.

All ellipticals in our sample show evidence of extended 
emission at 24$\mu$m similar to that of the optical stellar 
images -- details of this will be discussed in a separate publication. 
However, spatial extension is also evident at 70$\mu$m as shown in 
Table 1, but the emission at 160$\mu$m is not spatially resolved 
due to decreasing resolution at larger wavelengths.
Figure 10 shows the surface brightness at 70$\mu$m for 
eight galaxies together with the point response function (PRF) 
for the MIPS detector (including the first Airy ring).
All contours in Figure 10 are arbitrarily normalized to 50 at the center.
The (logarithmic) extension in arseconds beyond the PRF is best 
measured at 0.08 of the central fluxes, 
shown with arrows in Figure 10. 
The full width at 4/50 = 0.08 maximum 
(subsequently referred to as FW0.1M; last two columns in Table 1) 
of the PRF is 37$^{\prime\prime}$. 
Twelve galaxies in our full 
sample have FW0.1M greater than 40$^{\prime\prime}$, 
indicating that these galaxies are spatially resolved at 70$\mu$m.
When FW0.1M/2$R_e$ is plotted against $F_{70}/F_{160}$, 
there is no clear evidence that galaxies with more extended 
$F_{70}$ have colder dust. 

\section{A Simple Far-Infrared Spectral Energy Distribution}

We estimate the FIR spectral energy distribution (SED) 
from diffusely distributed interstellar 
dust in elliptical galaxies using a simple model similar 
to that described in Temi et al. (2003). 
For completeness and clarity, we briefly describe this model 
again here. 
We assume that dust grains
ejected from evolving stars come into direct contact 
with the hot interstellar gas after a short
time, $\lta 10^5-10^6$ yr, when the stellar gas has 
heated to the (virial) temperature of the hot interstellar gas. 
This time scale is established by comparing typical Balmer 
emission line observations with the luminosity expected 
if all the gas expelled from stars (at the expected
rate for an old stellar population) is ionized 
and in hydrostatic equilibrium with the hot gas 
(Mathews \& Brighenti 1999). 
After $\sim 10^6$ yrs this warm gas at $10^4$ K must merge with 
the hot gas since otherwise its Balmer line luminosity would 
exceed typically observed values. 
However, the dust in this gas can resist sputtering 
destruction in the hot 
interstellar gas ($T \sim 10^6 - 10^7$ K) for much longer times, 
$\sim10^8$ yrs. 
Consequently, we model ``recently'' produced dust that was 
created by local stars during the previous $\sim 10^8$ yrs.
Once in contact with the hot gas, the dust grains are destroyed 
by sputtering at a rate 
\begin{equation}
{da \over dt} = - 3.2 \times 10^{-14} n_p [ 1 +
(2 \times 10^6 / T)^{2.5} ]^{-1}
~~~\mu{\rm m}~{\rm s}^{-1},
\end{equation}
where $a$ is the grain radius (in $\mu$m) and $n_p$ is the proton 
density in the hot gas 
(Draine \& Salpeter 1979; Tsai \& Mathews 1995).

If $N(r,a,t)da$ be the number of grains cm$^{-3}$ with radius
between
$a$ and $a+da$ at galactic radius $r$ at time $t$, 
the grain population evolves according to
\begin{equation}
{\partial \over \partial a} \left[ N \left( {da \over dt}
\right) \right] = S(r,a),
\end{equation}
where $S(r,a)da$ is the rate
at which grains with radii $a$ are expelled from stars per cm$^3$ 
(Tsai \& Mathews 1995).
We assume $S(a)=S_o a^{-s}$ for $0<a<a_{\rm max}$, with $s=3.5$
(Mathis et al. 1977), so the coefficient 
\begin{equation}
S_o(r) = {3 \delta \alpha_* \rho_* \over
4 \pi \rho_g 10^{-12}} { (4 - s) \over a_{max}^{4-s}}
\end{equation}
depends on the rate that the old stellar
population ejects mass
$\alpha_* = M_*^{-1} |dM_*/dt| \approx 4.7 \times 10^{-20}$ s$^{-1}$
(Mathews, 1989),
the stellar density
$\rho_*$, the initial dust to gas
mass ratio scaled to the stellar metal abundance in solar units 
%in NGC4472,
$\delta = (1/150)z(r)$,
and the density of (amorphous) silicate grains,
$\rho_g = 3.3$ gm cm$^{-3}$. The solution of eq. (7) is 
\begin{equation}
N(r,a) = \left| {da \over dt} \right|^{-1}
{S_o(r) \over (s - 1) } a^{1-s}~~~a \leq a_{max}.
\end{equation}

The grain temperature $T_d$ 
is determined by the balance of heating (by both
absorption of stellar radiation and by electron-grain collisions)
and radiative cooling:
\begin{displaymath}
\int_0^{\infty} 4 \pi J_*(r,\lambda) Q_{abs}(a,\lambda)
\pi a^2 d \lambda
+ 4 \pi a^2 \case{1}{4} n_e \langle v_e E_e \rangle
\tau(a)
\end{displaymath}
\begin{equation}
= 4 \pi a^2 \sigma_{SB} T_d(r,a)^4
\langle Q_{abs} \rangle (T_d, a).
\end{equation}
In view of the high electron-grain collision frequency, 
we neglect stochastic heating due to single impacts.

We assume that all grains have properties similar to astronomical
silicates (Laor \& Draine 1993) using the following approximations
for the absorption coefficient $Q_{abs} \approx a \psi(\lambda)$,
and for the Planck-averaged values
$\langle Q_{abs} \rangle \approx 1.35 \times 10^{-5} T_d^2 a$
(see Draine \& Lee 1984).
The mean intensity of starlight
is found by integrating over a de Vaucouleurs
stellar profile using the 
appropriate effective radius $R_e$ and stellar luminosity 
for each galaxy. 
The collisional heating term in the equation above includes a correction
$\tau(a)$ for small grains that do not completely stop the
incident thermal electrons (Dwek 1986) and $\langle v_e E_e \rangle =
(32/\pi m_e)^{1/2} (k T)^{3/2}$ is an average over a Maxwellian
distribution.
The electron density and temperature of the hot gas are
calculated using deconvoluted profiles from X-ray observations. 
The FIR emissivity at each wavelength is calculated from 
\begin{displaymath}
j(r,\lambda) = {1 \over 4 \pi} \int_{0}^{a_{max}}
N(r,a) 4 \pi a^2 10^{-8} Q_{abs}(a,\lambda) 
\end{displaymath}
\begin{equation}
\times \pi B(T_d(r,a),\lambda) da.
\end{equation}

In Figure 11 we compare the expected FIR spectral energy distribution 
for three galaxies in our sample 
that have well-known hot gas density and temperature profiles 
from X-ray observations: 
NGC 1399, 1404 and 4472.
None of these galaxies are spatially resolved at 70$\mu$m.
The X-ray data for these galaxies has been taken from the following 
sources: 
NGC 1404 \& 1399: (Paolillo et al 2002; Jones et al. 1997); 
NGC 4472: (Mathews \& Brighenti 2003; Brighenti \& Mathews 1997).
The profiles in Figure 11 
show the expected FIR spectral energy distributions computed 
with the equations above.
The solid and dashed lines 
correspond respectively to emission from 
grains with original maximum sizes $a_{max} = 0.3\mu$m and 1.0$\mu$m.
The observed fluxes for NGC 1399 and NGC 4472 at 70$\mu$m and 160$\mu$m 
are in excellent agreement with our simple model.
The fit to NGC 1404 is also satisfactory, given the many assumptions 
in our model. 
While the flux at 24$\mu$m is also shown in Figure 11,
this circumstellar emission is not included in our model,
but it supports our assumption that much of the diffuse interstellar
dust in these three galaxies has a stellar origin.

It is of particular interest to determine 
FIR spectral energy distributions for galaxies that 
have accurate interstellar gas density and temperature 
profiles and are also spatially resolved at 70$\mu$m.
Seven galaxies in Table 1 have FW0.1M greater than
50$^{\prime\prime}$ but only six of these are in the
Ellis and O'Sullivan (2006) list of X-ray galaxies detected
with ROSAT: NGC 2768, 2974, 4125, 4636, 4696, and 5044.
Of these six galaxies, only three -- NGC 4636, 4696 and 5044 -- have 
X-ray luminosities
($\log L_x \gta 41.5$) large enough for an accurate determination
of the gas density and temperature profiles.
However, the optically peculiar elliptical NGC 4696 
is the central galaxy in the Centaurus Cluster 
with powerful FRI-type radio emission, a large mass of dusty neutral 
gas and highly disturbed hot cluster gas 
(Fabian et al. 2006). 
In view of these complexities, NGC 4696 is an inappropriate 
galaxy to learn about the nature of FIR emission 
from normal elliptical galaxies. 
We find that 
the observed fluxes for the two remaining FIR-extended galaxies 
-- NGC 5044 and NGC 4636 -- are 
very much larger than predicted by 
our steady state model for the FIR spectral energy distribution. 
We believe that the dust in these galaxies 
has an entirely different origin. 
The large and extended 
FIR emission from NGC 5044 and NGC 4636, which is related 
to the enormous range of FIR luminosities in 
Figure 3, will be discussed in a separate forthcoming publication.

\section{Discussion}

\subsection{Large Range of $L_{FIR}$}

The first important 
conclusion to be drawn from Figures 2 and  3 is that 
warm circumstellar dust and colder interstellar 
dust behave very differently.
The relatively small 
scatter in $F_{24}$ in Figure 2 is slightly larger than 
our estimated observational errors.
Nevertheless, we have been unable to find correlations 
between stellar age or metallicity 
and the residuals in $F_{24}$ from the mean correlation. 
We conclude that the warm dust emitting at 24$\mu$m is 
largely circumstellar and this is confirmed by the 
similarity of optical and 24$\mu$m surface brightness profiles 
which will be discussed in detail elsewhere. 

The second and most remarkable result in Figures 2 and 3 
is that the emission from colder dust at 70 and 160$\mu$m 
varies by a huge factor relative to that at 
24$\mu$m and optical wavelengths. 
In the previous section and in Figure 11 
we show that 70 and 160$\mu$m fluxes 
from three galaxies having well-observed interstellar gas density profiles 
can be fit reasonably well with our simple dust heating model 
without varying any of the dust parameters from those we assumed 
at the outset. 

However, notice that all  
three galaxies in Figure 11 that agree with our model SED 
-- NGC 1399, 1404 and 4472 -- have values of $L_{70}$ and $L_{160}$ 
in Table 2 
that lie near the bottom of the observed distributions in Figure 3.
We think it is impossible to explain the huge dispersion of 
$L_{70}$ and $L_{160}$ in Figure 3 by simply varying the 
dust parameters in 
the SED models used for the three galaxies in Figure 11,
i.e., the stellar mass loss rate, dust composition, 
dust size distribution, etc. are unlikely to be in error by 
an order of magnitude or more. 
Our SED model clearly fails to fit most of the FIR observations.

Some of the large spread of $L_{70}$ and
$L_{160}$ in Figure 3 can be understood because 
our sample includes very FIR-luminous elliptical galaxies that 
contain abnormal masses of interstellar
dust and cold gas, probably acquired by a recent merger. 
In Appendix A we review relevant observations of 12 sample galaxies 
having the highest 70$\mu$m luminosities in 
order of descending $L_{70}$. 
We conclude from this limited information that most ellipticals with the 
largest $L_{70}$ and $L_{160}$ in Figure 3 cannot be regarded 
as typical or normal. 
Of the 12 galaxies in Appendix A, 
we regard the following 8 as either likely 
S0s (which often have extended, rotationally supported, star-forming 
disks of cold dusty gas) 
or having massive HI or molecular lanes or disks
probably acquired by mergers:
IC 3370, NGC 5018, 2974, 4125, 5322, 2768, 3962, and 5077.
These atypical or mis-classified elliptical galaxies are the 
brightest galaxies in our sample. 
However, we regard the following galaxies in Appendix A 
with slightly lower $L_{70}$ 
as normal or quasi-normal ellipticals: 
NGC 3557, 5044, 4589, and (possibly) 4697.
If we disregard the eight galaxies with atypical dust content, 
the range in $L_{70}$ and $L_{160}$ occupied by normal 
ellipticals is reduced from $\sim 100$ 
(shown in Figure 3) to about $\sim 30$, 
but this is 
still larger than can be explained by varying dust parameters in our 
SED model. 
These normal ellipticals with excess FIR are 
the most interesting galaxies in our sample.  

\subsection{Location of FIR-Emitting Dust}

We assume that galaxies with $L_{FIR}$ significantly above
that predicted by our model SEDs in Figure 11 
have dust contents that exceed those that can be understood 
by a steady state balance between dust creation by 
mass-losing stars and dust destruction by sputtering. 
Galaxies in Figure 3 with $\log L_{70} \gta 40.5$ 
must contain additional cold dust from another source.
This additional cold dust 
is either (1) spread diffusely throughout a large region 
of the galactic hot interstellar gas or
(2) located in dense optically thick clouds similar to 
those seen in absorption by optical observers. 
In Appendix B we describe a simple argument 
that supports the first option above, diffuse FIR emission. 
We suppose in Appendix B that the FIR-emitting dust is 
in a small spherical cloud of radius $r_c$
located at the center of an elliptical galaxy. 
The dust is heated both by internal stars and by 
starlight incident from the entire galaxy. 
The dust cloud is assumed to be optically thick to most 
stellar radiation. 
As the cloud radius $r_c$ is increased, 
the estimated 70$\mu$m luminosity also increases because 
the cloud encloses more stars and intercepts more light 
from external stars. 
However, this simple dusty cloud model 
cannot account for the luminosity range in Figure 3 unless 
$r_c$ is so large ($\gta 200$ pc) 
that it would be easily apparent in moderate resolution optical images. 
The UV-ionized surfaces of these clouds may also be visible. 
While AGN heating within optically thick dusty clouds
could easily provide the needed FIR energy,
this seems unlikely since AGNs
would need to be bolometrically brighter than the stars within $r_c$
and would therefore
be seen in those ellipticals in which the dusty clouds are
in disks or not present at all.
Such luminous AGN would almost certainly
be visible at other frequencies.

Comparisons of the most FIR-luminous (normal) ellipticals in 
our sample with recent HST images 
(e.g. van Dokkum \& Franx 1995;
Lauer et al. 2005), 
show a correspondence between the absence of nuclear dust 
and low $L_{FIR}$.
For example, of the 19 galaxies in common between our sample 
and that of Lauer et al. (2005), 
all 11 of the ellipticals with $L_{70} > 10^{40.5}$ erg s$^{-1}$ 
have detectable nuclear dust, while all but one of the 9 galaxies with 
$L_{70} < 10^{40.5}$ have no nuclear dust detected. 
Nevertheless, even for these galaxies with lower $L_{70}$ 
we believe that most 
of the $L_{FIR}$ comes from a diffuse component. 
In addition to the argument in Appendix B, 
our preference for extended excess dust is supported by noting that  
all six galaxies with extended FIR emission in Table 1
(e.g. FW0.1M $> 50^{\prime\prime}$) have excess FIR emission, 
i.e. $L_{70} > 10^{40.5}$ erg s$^{-1}$. 

\subsection{External Origin for the Excess Dust}

If the FIR emission in excess of that expected 
from our simple model SED in Figure 11 is from cold dust diffusely 
distributed throughout the 
hot interstellar gas, where has it come from? 
It is often supposed that optically-observed dust in elliptical galaxies 
has an external origin, resulting from 
a recent merger with a small dust-rich galaxy that is otherwise
difficult to observe. 
We adopted this explanation previously 
in our discussion of the excess dust in NGC 4636 
(Temi, et al. 2003), which we interpreted as a dusty merger that 
occurred within the last $\sim 3 \times 10^8$ yrs, the estimated
sputtering time in the hot gas in this galaxy. 
However, it has since been determined 
that the mean stellar age in NGC 4636 
is very old (10.3 Gyrs, Sanchez-Blazquez et al. 2006), 
and this strongly argues against a merger $\sim 3 \times 10^8$
yrs ago with a dusty, star-forming galaxy. 
Moreover, we observe that the additional dust in NGC 4636 is extended 
(FW0.1M = 54.4$^{\prime\prime}$ in Table 1) and appears to be 
distributed rather evenly about the center unlike the 
irregular distribution that might be expected 
to result from a recent merger undergoing dynamical friction 
with accompanying ram-stripping and tidal disruption.
So our previous interpretation of NGC 4636 needs to be reevaluated.

The dust sputtering time for grains of initial radius $a$  
in the hot gas of an elliptical galaxy is 
$t_{sp} \approx a / |da/dt|_{sput} \approx 1.2 \times 10^8 a_{\mu}
(n_e/10^{-2})^{-1}$ yrs (where $a_{\mu}$ is in microns) 
provided the ambient gas temperature is $T \gta 10^6$ K.
(Draine \& Salpeter 1979; Tsai \& Mathews 1995). 
Even if the dust is mixed with warm gas at $T \sim 10^4$ K, 
where sputtering is negligible, 
its overall sputtering destruction time $t_{sp}$ 
cannot be significantly reduced since,
as discussed below, we generally expect that the warm gas 
is thermally heated and joins the (spatially extended) 
hot gas phase on a time scale
$\sim 10^6$ yrs that is very much less than $t_{sp}$. 
Therefore it is reasonable to assume that the diffuse FIR-emitting dust 
in our sample galaxies is in direct contact with the hot gas 
during most of its sputtering lifetime. 

The dynamical (freefall) time of a merging dusty galaxy across 
the optical image of a large elliptical is several $\sim 10^8$ years, 
comparable to $t_{sp}$. 
Therefore, on the merger hypothesis, diffusely distributed 
cold dust must be resupplied by mergers in a regular 
fashion at least every $\sim 10^8$ yrs.
Therefore, an entourage of 
dusty merging galaxies should be commonly observed 
in the outer regions of the majority of 
elliptical galaxies having FIR excess emission -- but they are not. 
In fact, mergers between elliptical and (small) dust rich galaxies
are rarely observed at zero redshift. 
While we believe that mergers do occur and that some of the most 
FIR luminous galaxies in our sample can only be explained by 
mergers, the similarity of 
the sputtering and dynamical time scales makes 
it difficult to accept the proposition that 
mergers are the dominant or 
only source of (extended) dust in most luminous 
elliptical galaxies. 
This point of view is supported by Figure 5, showing that the 
incidence of young stars in elliptical galaxies is uncorrelated 
with the FIR excess luminosities seen in Figure 3.

\subsection{Internal Origin for Excess Dust}

\subsubsection{Evolution of Dust Ejected from Stars}

Before exploring the origin of the additional dust 
required to produce FIR luminosities in excess of 
$L_{70} \sim 10^{40.5}$ erg s$^{-1}$, it is useful to review 
the evolution of cold dusty gas that is introduced into the hot, 
X-ray emitting interstellar gas by mass-losing red giant stars. 
Our SED model is based on an internal stellar source of dust which 
is confirmed by mid-IR observations 
of (apparently outflowing) circumstellar 
dust (Knapp et al. 1989; Athey et al. 2002;
Bressan et al. 2006; Bregman et al. 2006).
We think that the fate of this dusty gas 
varies with galactic radius.

Throughout most of the volume occupied by galactic stars beyond 
$\sim 1$ kpc, we assume that both the dust and gas ejected from 
red giant stars ultimately merge with the hot gas which is the 
dominant interstellar phase. 
We expect that gas recently ejected from stars
can be maintained in pressure equilibrium at warm temperatures
($T \approx 10^4$ K) by photoionization by UV emitted by
post-AGB stars.
Emission from warm gas at $r \gg 1$ kpc is faint and has 
not been observed.
Eventually this warm gas thermally mixes with nearby hot gas. 
When the dust in this gas confronts the hot gas 
it will be destroyed by sputtering in $\sim 10^8$ yrs. 
 
However, within $\sim$1 kpc of the centers of elliptical 
galaxies the evolution of dusty stellar ejecta is different. 
Extended warm gas in this region is commonly observed, but 
its velocities differ from those of nearby stars 
(e.g. Caon, Macchetto \& Pastoriza 2000).
Evidently the hot and warm gas exchanged momentum when 
the warm gas had a much lower density. 
After dusty stellar envelopes and winds are ejected 
from orbiting red giant stars, 
the gas expands until its density approaches (but still exceeds)
the local hot gas density. 
At this time the ejected gas, which is still moving with the 
stellar orbital velocity, is disrupted by hydrodynamic
instabilities and shocked to a few million degrees 
where its density is comparable to that of the hot interstellar gas. 
During this period of interaction between the stellar ejecta 
and the hot gas, momentum is exchanged and the 
ejected gas acquires some of the velocity of the ambient hot gas. 
Even if the gas ejected from stars is (transiently) 
heated completely to the temperature of the local hot gas 
($\sim10^7$ K), its dust content will cool the gas rapidly 
back to $\sim10^4$ where it is stabilzed by UV from post-AGB
stars.
This rapid dust-assisted cooling occurs because 
of very frequent inelastic collisions between thermal 
electrons in the hot gas and the dust grains. 
Mathews \& Brighenti (2003) showed that
efficient dust cooling of the hot gas is also faster than
the sputtering lifetime of the dust, so the gas retains
some dust after it cools to $\sim 10^4$ K.
Since the cooling time to $\sim 10^4$ K within 1 kpc 
is less than the dynamical time in the local (stellar) 
potential,  
these dense warm clouds freefall toward the galactic 
center and accumulate there.  
As dusty warm gas collects into a disk near the center, 
similar to those observed by 
van Dokkum \& Franx (1995) and Lauer et al. (2005), 
it eventually becomes optically thick to the stellar UV radiation. 
Because of its residual dust, we expect 
molecular formation and eventual cooling far below $10^4$ K.  
Dense, cold nuclear dust disks and clouds similar to those observed 
can form naturally from stellar mass loss in elliptical galaxy cores 
(Mathews \& Brighenti 2003).
Dusty galaxy mergers are not required. 
We speculate below that this central dust reservoir may be the 
source of extended diffuse FIR emission in many ellipticals. 

Central dusty clouds and disks in ellipticals are not merely
an amusing curiosity, but are likely to be key to understanding
the energetics of AGN feedback in regulating cooling flows 
in the hot gas within and around elliptical galaxies.
Indeed, Simoes Lopes et al. (2006) find a correlation between
nuclear dust and AGN activity in elliptical galaxies. 
Consider the black hole accretion
luminosity expected in a large group-centered elliptical galaxy
by dusty mass loss just within the central kpc where
rapid dust-assisted cooling is expected.
Accretion is expected to quickly follow cooling.
The AGN luminosity is $L_{agn} = \epsilon {\dot M} c^2
= 6 \times 10^{45} [{\dot M}_1/(M_{\odot}~{\rm yr}^{-1})]
(\epsilon / 0.1)$ erg yr$^{-1}$ where ${\dot M}_1$ is the
total rate of stellar mass loss within the central kpc.
Since a $10^{10}$ year old stellar population of mass $M_*$ ejects
$1.5 (M_*/10^{12}M_{\odot})$ $M_{\odot}$ yr$^{-1}$,
the stellar mass within 1 kpc, $\sim 5 \times 10^{10}$ $M_{\odot}$,
generates ${\dot M}_1 \approx 0.07$ $M_{\odot}$ yr$^{-1}$.
When accreted by the central massive black hole,
this translates into a luminosity $L_{agn} \approx 4 \times 10^{44}$
ergs s$^{-1}$ that exceeds the entire X-ray luminosity
of the most luminous galaxy groups, $L_x \sim 3 \times 10^{43}$
ergs s$^{-1}$.
Consequently, dusty stellar mass loss within the central kpc of
bright ellipticals can generate enough accretion energy
to balance the entire bolometric X-ray luminosity of a large
galaxy group.
The heating of the hot gas as a consequence of this accretion 
also provides enough energy to drive buoyant
mass circulation in the hot gas out to 5-10 kpc.

The evolution of dusty disks in galactic cores that 
we just described cannot continue very long. 
If small core disks
accrete dusty gas at ${\dot M}_1 \approx 0.07$ $M_{\odot}$ yr$^{-1}$,
for $\sim 10^{10}$ yrs, and allowing for a larger mass-loss
rate in the past, currently observed
disks should have a typical mass of $\sim 10^{10}$ $M_{\odot}$
and a dust
mass of $\sim 10^{8}$, both far in excess of those observed.
(Small stellar disks with masses $\sim 10^9 - 10^{10}$ $M_{\odot}$
are not commonly observed in the cores of most ellipticals,
so normal star-formation is ruled out.)
A large minority of bright ellipticals show no evidence of
central dusty disks/clouds.
Where has this dusty gas gone?

\subsubsection{Explosive Ejection of Dust from the Galactic Core}

The excess diffuse dust observed in FIR-emission
may have been exploded out from the centers of elliptical galaxies
by the release of AGN energy.
We estimate from our model SED 
that the total mass of excess dust required to 
explain the FIR luminosities in Figure 3 is 
$M_d \sim 3 \times 10^5$ 
$M_{\odot}$. 
This is comparable to the estimated dust masses 
in the central dusty disks or lanes 
(e.g. Goudfrooij \& de Jong  1995). 
If the (very uncertain) dust to gas mass 
ratio is $\delta \approx 0.01$, then 
the mass of gas that initially contained this dust
is $M \sim 10^7 - 10^8$ $M_{\odot}$.
If cold dusty gas is ballistically 
ejected from the core out to a few kpc,
velocities of $\sim100$ km s$^{-1}$ are required.
The kinetic energy needed to explosively transport
dusty gas from 
the galactic core to $\sim1$ kpc is $E_k \approx 0.5 M v^2
= 10^{55} (0.01/\delta)(M_d/10^{6}~M_{\odot}) (v/100{\rm km/s})^2$
ergs.
Although very uncertain,
the energy involved is certainly larger than that of a
single supernova. 
Since multiple starburst activity in elliptical galaxies is 
very rare, 
AGN energy is required for dust ejection from the core. 
If the AGN dust ejection process is intermittent, this would explain 
those elliptical galaxies that are not observed to have 
optically obscuring dust in their cores 
or extended FIR excess emission. 
The FIR variance expected from intermittent AGN 
outbursts obviates the argument 
of Forbes (1991) and Simoes Lopes et al. (2006) 
that the failure of an $L_{FIR}$-$L_B$ correlation 
necessarily requires an external source of dust by mergers. 
Some explosive character to the AGN-dust cloud interaction is 
implied by the chaotic nature of the central dust observed 
in many HST images (Lauer et al. 2005). 
This chaotic, 
denser dusty gas is highly transient and must have been ejected 
from the core. 
These chaotic clouds cannot be understood as infalling 
stellar ejecta (as suggested by Lauer et al.), 
formed by the coalescence of many dusty stellar
envelopes produced by individual stars, since the 
free fall time ($\sim 10^6$ yrs) to the galaxy center 
is too short for such accumulations to occur. 

However, explosive ejections of central dust to $\sim 5 - 10$ kpc 
with energies $E_k \gta 10^{55}$ ergs 
would produce kpc-sized central cavities, strong shocks and heated 
cores that are rarely observed in X-ray images 
while extended excess FIR emission is common. 
This may be the best argument against explosive dust transport. 

\subsubsection{Buoyant Ejection of Dust form the Galactic Core}

Density irregularities and multi-temperature X-ray spectra 
of the hot interstellar gas in elliptical galaxies reveal 
buoyant activity, presumably related to a less explosive 
heating of the hot gas with AGN energy 
(Buote, Lewis, Brighenti \& Mathews 2003a,b). 
In view of these X-ray observations, 
it is plausible that central dust can be buoyantly transported
out to $\sim 5-10$ kpc, producing the extended FIR emission 
that we observe.

The buoyant rise time $t_{buoy}$ of heated gas is 
expected to be comparable to or 
somewhat less than the sputtering destruction time 
in the buoyant gas.
For example in the galaxy/group NGC 5044 the freefall time 
from radius $r_{kpc}$ to the galactic center is 
$t_{ff} = 4.5 \times 10^6 r_{kpc}^{0.87}$ yrs, 
based on a mass model 
that combines a de Vaucouleurs stellar profile 
($M_* = 3.37 \times 10^{11}$, 
$M_{\odot}$; $r_{e} = 10$ kpc) with an Navarro-Frenk-White 
dark halo 
(mass $M_{halo} = 4 \times 10^{13}$ $M_{\odot}$; concentration
$c = 8.67$).
The buoyancy rise time $t_{buoy} = \beta t_{ff}$ is greater than 
$t_{ff}$ by some factor $\beta \gta 1$. 
>From Draine \& Salpeter (1979) the 
approximate grain sputtering time 
is $t_{sput} \approx 2.4 \times 10^6 a_{\mu} n_e(r)^{-1}$ 
where $a_{\mu}$ is the grain radius in microns.
Adopting $n_e \approx 0.085
r_{kpc}^{-0.87}$ cm$^{-3}$ for the observed hot gas 
density in NGC 5044 within 30 kpc, 
the local sputtering time 
is nearly proportional to $t_{ff}$,
$t_{sput} \approx 1.8 a_{\mu}t_{ff}$.
However, the buoyant gas will have a density that is 
$\delta < 1$ times lower than the observed hot gas density.
Therefore, we see that the ratio of 
the local sputtering time in the buoyant gas to the 
buoyancy rise time is $\sim 1.8a_{\mu}/(\delta \beta)$, 
which may well exceed unity  
considering the many uncertainties involved. 

The exact nature of the AGN energy release 
that initiates buoyant activity is unclear at present. 
In about half of the ellipticals with
optically visible dust, the images of the obscuring dust are
morphologically ``chaotic'' and distributed
in an asymmetric fashion away from the center of the galaxy
(Lauer et al. 2005).
Evidently the AGN energy is sufficient to disrupt the dense 
central dusty disks and lanes which must reform again 
later by local stellar mass loss. 
It is interesting that the radii of the dusty disks observed by 
Lauer et al. (2005) do not correlate with the 
stellar rotation, also suggesting short disk lifetimes and
intermittent disruption of the dusty disks.

Regardless of the mode of outward dust transport, explosive 
or buoyant, when the dust rises to 5-10 kpc it may 
cause the hot gas to cool back to 
warm gas temperatures $\sim 10^4$ K. 
The correlations in Figure 8 between $F_{70}$ and $F_{160}$ 
and the optical line fluxes, if real, indicate a connection 
between the warm gas and (excess) dust. 
Similar morphologies of gas and dust images have been discussed
by optical observers 
(e.g. Goudfrooij et al. 1994).  
Although optical line fluxes appear 
to correlate with the FIR and B-band fluxes, 
Figure 2 shows that the FIR and B-band fluxes do not 
mutually correlate. 
This can be understood if most of the FIR-emitting dust comes
not from recent stellar mass loss, but is transported out from 
a reservoir in the galactic core, 
which contains a large mass of dust unrelated  
to the the global stellar B-band flux or luminosity. 

\section{Conclusions}

Our main conclusions are:

\vskip.1in
\noindent
(1) The optical luminosities $L_B$ of elliptical 
galaxies correlate with luminosities at 24$\mu$m ($L_{24}$) but not
with $L_{70}$ or $L_{160}$.
This indicates that the transition between circumstellar and interstellar 
dust occurs for dust grains with emission that peaks somewhere 
between 24$\mu$ and 70$\mu$m.

\vskip.1in
\noindent
(2) Elliptical galaxies with similar $L_B$ have $L_{70}$ and $L_{160}$
that vary over a factor $\sim 100$. 
This variation is reduced to $\sim 30$ if abnormally dust-rich
ellipticals are excluded from the sample. 

\vskip.1in
\noindent
(3) Neither $L_{70}/L_B$ nor $L_{160}/L_B$ correlates with the 
stellar age or metallicity of the galaxies. 

\vskip.1in
\noindent
(4) Weak correlations appear between optical line emission 
fluxes from warm ($\sim 10^4$ K) gas and 
$F_{70}$ and $F_{160}$. 
This suggests a relationship between 
dust and gas that may have cooled from the hot phase.

\vskip.1in
\noindent
(5) Seven galaxies in our sample are clearly extended 
at 70$\mu$ with FW0.1M exceeding 50$^{\prime\prime}$ 
corresponding to 5--10 kpc.

\vskip.1in
\noindent
(6) The luminosities and spectral energy distributions 
of elliptical galaxies with the lowest $L_{70}$ and $L_{160}$
are consistent with a simple steady state dust model 
in which dust expelled from stars is eroded by sputtering 
in the hot interstellar gas. 

\vskip.1in
\noindent 
(7) We show in Appendix B that the observed emission at 
70 and 160$\mu$m cannot come from small optically visible 
clouds in the galactic cores similar to those seen in 
HST images.

\vskip.1in
\noindent
(8) In view of the common occurrance of 
spatially extended emission at 70$\mu$m 
in many ellipticals, it is unlikely that this dust 
can be supplied by successive mergers with gas-rich galaxies, 
each occurring during the relatively short 
sputtering lifetime of the diffuse dust, $\sim 10^8$ yrs. 
Addiiontal supporting evidence for such mergers is lacking. 

\vskip.1in
\noindent
(9) The appearance of extended FIR-emitting dust in 
otherwise normal elliptical galaxies can be 
understood by the disruption and 
buoyant transport of 
centrally accumulated dust by intermittent AGN activity. 
The buoyant rise time out to 5-10 kpc, $\sim 10^7$ yrs, 
is less than the sputtering lifetime of the dust, $\sim 10^8$ yrs.

\acknowledgements
This work is based on observations made with the Spitzer Space 
Telescope, which is operated by the Jet Propulsion Laboratory, 
California Institute of Technology, under NASA contract 1407. 
Support for this work was provided by NASA through Spitzer 
Guest Observer grant RSA 1276023.
Studies of the evolution of hot gas in elliptical galaxies
at UC Santa Cruz are supported by
NASA grants NAG 5-8409 \& ATP02-0122-0079 and NSF grant
AST-0098351 for which we are very grateful. 
We thank Justin Harker for his help with the photometric 
properties of old stellar populations.

%\end{document}

\clearpage

\begin{deluxetable}{lccrrrrrrrrrr}
\tabletypesize{\scriptsize}
%\rotate
%\tablecaption{MIPS Far-infrared photometry }
\tablecaption{}
\tablewidth{0pt}
\tablehead{
\colhead{Name} & \colhead{Sample\tablenotemark{a}} & \colhead{T\tablenotemark{b}}&\multicolumn{3}{c}{Flux Density} &\colhead{D\tablenotemark{c}} & \colhead{$R_e$} & \multicolumn{2}{c}{70$\mu$m FWHM} &\colhead{}& \multicolumn{2}{c}{ 70$\mu$m FW0.1M} \vspace{0.1cm}\\ 
\cline{4-6}\cline{9-10}\cline{12-13}\\ 
\colhead{} & \colhead{} & \colhead{} & 24 $\mu$m & 70 $\mu$m & 160 $\mu$m & &  &           &           & \colhead{}  &           &           \vspace{0.1cm}\\
 &(PI) & & (mJy) & (mJy) & (mJy) & (Mpc) & ($\prime \prime$) &  ($\prime \prime$) & (Kpc) & & ($\prime \prime$) & (Kpc)
}
\startdata
NGC~507  & Temi      &-3.3&\nodata      & $< 2.8$      &\nodata       &71.99& 77.1 &      &     &  &           &             \\
NGC~533  & Temi      &-4.8& 7.1$\pm$3.1 & 23.7$\pm$4.9 &26.1$\pm$5.6  &68.23& 47.5 &      &     &  &           &             \\
NGC~584  & Kennicutt &-4.6&47.6$\pm$7.6 &49.8$\pm$6.9  & $< 17.2$     &23.76& 27.4 &      &     &  &           &             \\
NGC~596  & Surace    &-4.3&15.9$\pm$4.6 &21.4$\pm$5.1  &16.7$\pm$7.3  &23.87& 30.0 &      &     &  &           &             \\
NGC~636  & Surace    &-4.8&10.3$\pm$3.6 &$< 16.4$      & $< 24.5$     &22.28& 18.9 &      &     &  &           &             \\
NGC~720  & Temi      &-4.8&25.6$\pm$5.1 &18.2$\pm$4.9  & $< 21.0 $    &22.29& 39.5 &      &     &  &           &             \\
NGC~855  & Kennicutt &-4.6&72.3$\pm$8.5 &1335$\pm$36   &2086$\pm$46   &9.33 & 18.9 & 22.5 & 1.0 &  & 58.2      & 7.6         \\
NGC~1377 & Kennicutt &-2.1&1603$\pm$40  &4988$\pm$71   &2375$\pm$48   &22.20&      &      &     &  &           &             \\
NGC~1395 & Kaneda    &-4.8&45.4$\pm$6.5 &129$\pm$11    &194$\pm$14    &21.98& 45.4 &      &     &  &           &             \\
NGC~1399 & Temi      &-4.5&59.9$\pm$7.3 &15.9$\pm$7.6  &24.1$\pm$8.6  &19.40& 42.4 &      &     &  &           &             \\
NGC~1404 & Kennicutt &-4.7&56.6$\pm$6.3 &32.6$\pm$6.3  & $< 32.2$     &19.40& 26.7 &      &     &  &           &             \\
NGC~1407 & Kaneda    &-4.5&42.5$\pm$6.4 &\nodata       & $< 35.4$     &22.08& 71.8 &      &     &  &           &             \\
NGC~1700 & Surace    &-4.7&18.7$\pm$4.9 &29.3$\pm$5.3  &40.3$\pm$7.2  &38.04& 13.7 &      &     &  &           &             \\
NGC~2325 & Temi      &-4.7&27.0$\pm$6.4 &36.5$\pm$6.8  &21.7$\pm$6.4  &31.92& 86.5 &      &     &  &           &             \\
NGC~2768 & Temi      &-4.3&46.6$\pm$8.2 &694 $\pm$26   &377 $\pm$29   &22.38& 53.5 & 21.4 & 2.3 &  & 56.4      & 6.1         \\
NGC~2974 & Kaneda    &-4.7&62.5$\pm$12.4&682 $\pm$18   &1979$\pm$54   &30.33& 36.9 & 22.8 & 3.4 &  & 55.4      & 8.1         \\
NGC~2986 & Temi      &-4.7&13.7$\pm$4.8 &$< 2.8$       & $< 14.5$     &32.51& 41.4 &      &     &  &           &             \\
NGC~3265 & Kennicutt &-4.8&287 $\pm$17  &1952$\pm$44   &2312$\pm$49   &21.33&      &      &     &  &           &             \\
NGC~3379 & Fazio     &-4.8&65.5$\pm$8.8 &60.5$\pm$8.0  &59.1$\pm$8.2  &10.71& 35.2 &      &     &  &           &             \\
NGC~3557 & Temi      &-4.8&30.5$\pm$4.5 &263 $\pm$16   &247 $\pm$18   &34.51& 37.8 & 20.0 & 3.3 &  & 38.8      & 6.5         \\
NGC~3610 & Surace    &-4.2&18.1$\pm$4.5 &19.1$\pm$7.3  &32.8$\pm$16.5 &29.24& 12.8 &      &     &  &           &             \\
NGC~3923 & Temi      &-4.6&42.9$\pm$6.7 &22.7$\pm$5.4  &44.0$\pm$7.8  &19.14& 53.3 &      &     &  &           &             \\
NGC~3962 & Kaneda    &-4.7&21.5$\pm$5.6 &374 $\pm$18   &495 $\pm$24   &23.23& 34.4 & 22.6 & 2.5 &  & 44.0      & 5.0         \\
NGC~4125 & Kennicutt &-4.8&53.7$\pm$6.9 &773 $\pm$28   &1395$\pm$37   &27.79& 59.9 & 21.6 & 2.9 &  & 56.8      & 7.7         \\
NGC~4365 & Cote'     &-4.8&21.7$\pm$4.7 &63.8$\pm$8.0  &52.9$\pm$7.3  &17.06& 57.2 &      &     &  &           &             \\
NGC~4472 & Fazio     &-4.8&73.3$\pm$8.6 &58.1$\pm$7.6  &60.4$\pm$7.8  &17.06&104.0 &      &     &  &           &             \\
NGC~4552 & Kennicutt &-4.6&57.3$\pm$7.8 &92.1$\pm$10.2 &171 $\pm$16   &17.06& 30.0 &      &     &  &           &             \\
NGC~4589 & Kaneda    &-4.8&14.9$\pm$4.5 &262 $\pm$16   &360 $\pm$21   &26.30& 41.4 & 22.6 & 2.9 &  & 45.6      & 5.8         \\
NGC~4621 & Cote'     &-4.8&34.2$\pm$6.3 &31.7$\pm$5.7  &43.4$\pm$6.6  &17.06& 46.5 &      &     &  &           &             \\
NGC~4636 & Fazio     &-4.8&31.2$\pm$5.6 &188 $\pm$12   &169 $\pm$14   &17.06&101.7 & 22.3 & 1.8 &  & 54.4      & 4.5         \\
NGC~4649 & Fazio     &-4.6&106 $\pm$10  &46.3$\pm$6.8  & $< 27.5$     &17.06& 73.6 &      &     &  &           &             \\
NGC~4660 & Cote'     &-4.7&15.2$\pm$4.3 &36.4$\pm$6.2  &53.7$\pm$8.2  &17.06& 12.8 & 20.0 & 1.7 &  & 39.0      & 3.2         \\
NGC~4696 & Sparks    &-3.7&22.9$\pm$4.7 &127 $\pm$13   &269 $\pm$21   &39.65&212.4 & 22.6 & 4.4 &  & 52.2      & 10.1        \\
NGC~4697 & Surace    &-4.8&43.8$\pm$6.7 &589 $\pm$25   &755 $\pm$28   &16.22& 75.3 &      &     &  &           &             \\
NGC~4915 & Surace    &-4.5&11.5$\pm$5.5 &29.2$\pm$4.8  &54.2$\pm$8.9  &46.98& 7.9  &      &     &  &           &             \\
NGC~5018 & Surace    &-4.6&60.5$\pm$9.1 &1119$\pm$34   &1687$\pm$41   &32.36& 24.9 & 20.4 & 3.2 &  & 39.6      & 6.21        \\
NGC~5044 & Temi      &-4.7&23.0$\pm$5.3 & 230$\pm$14   & 242$\pm$18   &32.36& 78.6 & 23.1 & 3.6 &  & 53.6      & 8.4         \\
NGC~5077 & Fazio     &-4.8&32.4$\pm$5.7 &127 $\pm$11   &141 $\pm$13   &32.36& 25.0 & 20.2 & 3.2 &  & 49.5      & 7.8         \\
NGC~5322 & Fazio     &-4.8&41.4$\pm$6.4 &455 $\pm$23   &625 $\pm$27   &29.79& 35.2 &      &     &  &           &             \\
NGC~5557 & Surace    &-4.8&13.2$\pm$5.0 &25.1$\pm$10.5 & $< 42  $     &49.02& 28.6 &      &     &  &           &             \\
NGC~5813 & Fazio     &-4.8&14.9$\pm$4.9 &58.4$\pm$7.6  &34.8$\pm$5.9  &30.78& 48.6 & 19.6 & 2.9 &  & 49.8      & 7.4         \\
NGC~5846 & Fazio     &-4.7&38.5$\pm$6.2 &102 $\pm$10   &117 $\pm$12   &24.55& 82.6 & 20.4 & 2.4 &  & 43.1      & 5.1         \\
NGC~5982 & Surace    &-4.8&13.5$\pm$5.1 &37.2$\pm$7.2  &59.9$\pm$11.7 &40.18& 29.4 &      &     &  &           &             \\
NGC~6703 & Fazio     &-2.8&19.9$\pm$4.5 &41.4$\pm$6.4  &17.9$\pm$5.1  &32.06& 23.8 &      &     &  &           &             \\
NGC~7619 & Temi      &-4.7& 9.9$\pm$4.7 &$< 3.8$       & $< 14.5$     &42.85& 32.1 &      &     &  &           &             \\
IC~3370  & Kaneda    &-4.7&43.8$\pm$6.7 &649 $\pm$25   &873 $\pm$30   &45.90& 18.5 &  20.5& 4.5 &  & 39.7      & 8.8         \\

\enddata
\tablenotetext{a}{Galaxies are selected from the following Spitzer observing programs:
SINGS Legacy Program, R. Kennicutt (PI);
GTO program number 69, G. Fazio (PI);
GO program number 3403, J. Surace (PI);
GO program number 3506, Sparks (PI);
GO program number 3619, H. Kaneda (PI);
GO program number 3649, P. Cote' (PI);
GO program number 20171, P. Temi (PI).
}
\tablenotetext{b}{The morphological type T is taken from the HyperLeda database.}
\tablenotetext{c}{Distances are calculated with $H_0=70$ km s$^{-1}$ Mpc$^{-1}$.}

\end{deluxetable}

\begin{deluxetable}{lrrrrrrr}
\tabletypesize{\scriptsize}
%\rotate
\tablecaption{}
\tablewidth{0pt}
\tablehead{
\colhead{Name} &  \colhead{Log $L_{24}$\tablenotemark{a}} & \colhead{Log $L_{70}$\tablenotemark{a}} &\colhead{Log $L_{160}$\tablenotemark{a}} & \colhead{Log $L_B$\tablenotemark{a}} & \colhead{Log $L_X$\tablenotemark{a}} & \colhead{age\tablenotemark{b}} &  \colhead{Z\tablenotemark{b}} \\
               &  \colhead{(erg/sec)}& \colhead{(erg/sec)}& \colhead{(erg/sec)}&\colhead{($L_{B,\odot}$)}&\colhead{(erg/sec)}&\colhead{(Gyr)}&\colhead{(dex)} 
}
\startdata
     NGC~507  &\nodata  &$<40.30 $& \nodata &  11.02  &  42.76  &    8.1$\pm$1.9 \  &  0.261$\pm$0.071 \ \\
     NGC~533  &  41.04  & 41.19   &  40.76  &  10.96  &  42.29  & \nodata &\nodata \\
     NGC~584  &  40.95  &  40.59  &$<39.66$ &  10.42  &$<40.15$ &    2.8$\pm$0.3 \ &  0.478$\pm$0.046 \ \\
     NGC~596  &  40.48  &  40.23  &  39.65  &  10.27  &$<39.66$ &    2.8$\pm$0.4 \ &   0.360$\pm$0.035 \ \\
     NGC~636  &  40.29  &$<40.11$ &$<39.82$ &  10.06  &$<40.17$ &    4.4$\pm$0.6 \ &  0.376$\pm$0.028 \ \\
     NGC~720  &  40.62  &  40.10  &$<39.70$ &  10.44  &  40.67  &    5.4$\pm$2.4 \ &  0.485$\pm$0.083 \ \\
     NGC~855  &  40.38  &  41.27  &  41.00  &  8.95   &$<39.83$ & \nodata & \nodata\\
     NGC~1377 &  42.48  &  42.59  &  41.81  &  9.60   &\nodata  & \nodata & \nodata\\
     NGC~1395  &  40.86  &  40.94  &  40.65  &  10.50  &  40.95  &    7.6$\pm$1.4 \ &  0.439$\pm$0.033 \ \\
     NGC~1399  &  40.87  &  39.92  &  39.64  &  10.58  &  41.69  & \nodata &\nodata \\
     NGC~1404  &  40.85  &  40.56  &$<39.76$ &  10.41  &  41.25  & \nodata &\nodata \\
     NGC~1407  &  40.83  &$<28.83$ &$<39.91$ &  10.64  &  41.06  &    7.4$\pm$1.8 \ &  0.379$\pm$0.044 \ \\
     NGC~1700  &  40.95  &  40.77  &  40.44  &  10.55  &  40.66  &    2.6$\pm$0.3 \ & 0.500$\pm$0.049 \ \\
     NGC~2325  &  40.96  &  40.71  &  40.01  &  10.66  &  40.76  & \nodata &\nodata \\
     NGC~2768  &  40.88  &  41.68  &  40.95  &  10.63  &  40.44  &     10.0$\pm$7.0\tablenotemark{c}  &   0.14$\pm$0.25\tablenotemark{c}\\
     NGC~2974  &  41.27  &  41.94  &  41.94  &  10.56  &  40.64  & \nodata &\nodata \\
     NGC~2986  &  40.68  &$<39.61$ &$<39.86$ &  10.57  &  41.02  &   11.2$\pm$1.9 \ &  0.347$\pm$0.045 \ \\
     NGC~3265 &  41.69  &  42.15  &  41.76  &   9.46  & \nodata & \nodata & \nodata\\
     NGC~3379  &  40.39  &  39.98  &  39.51  &  10.12  &$<39.60$ &    10.0$\pm$1.1 \ &  0.299$\pm$0.036 \ \\
     NGC~3557  &  41.08  &  41.64  &  41.15  &  10.82  &  40.64  & \nodata &\nodata \\
     NGC~3610  &  40.71  &  40.36  &  40.13  &  10.46  &  39.89  &    1.7$\pm$0.1\tablenotemark{c}  &   0.76$\pm$0.16\tablenotemark{c}\\
     NGC~3923  &  40.71  &  40.06  &  39.88  &  10.58  &  40.72  &    3.3$\pm$0.8 \ &  0.623$\pm$0.070 \ \\
     NGC~3962  &  40.58  &  41.45  &  41.10  &  10.34  &$<40.28$ &\nodata  & \nodata \\
     NGC~4125  &  41.14  &  41.92  &  41.71  &  10.86  &  41.0   & \nodata &\nodata \\
     NGC~4365  &  40.32  &  40.41  &  39.86  &  10.40  &  40.31  &   7.9$\pm$1.2\tablenotemark{d}  &  0.344$\pm$0.065\tablenotemark{d} \\
     NGC~4472  &  40.85  &  40.37  &  39.92  &  10.96  &  41.49  &    9.6$\pm$1.4  \ &  0.342$\pm$0.046 \ \\
     NGC~4552  &  40.74  &  40.57  &  40.37  &  10.35  &  40.77  &   12.4$\pm$1.5  \ &  0.356$\pm$0.034 \ \\
     NGC~4589  &  40.53  &  41.40  &  41.07  &  10.39  &  40.42  & \nodata & \nodata \\
     NGC~4621  &  40.52  &  40.11  &  39.78  &  10.38  &  40.08  &   8.8$\pm$1.1\tablenotemark{d}  &  0.299$\pm$0.050\tablenotemark{d} \\
     NGC~4636  &  40.48  &  40.88  &  40.37  &  10.57  &  41.65  &   10.3$\pm$1.3\tablenotemark{d}  &  0.181$\pm$0.104\tablenotemark{d}\\
     NGC~4649  &  41.01  &  40.27  &$<39.57$ &  10.79  &  41.34  &   14.1$\pm$1.5  \ &  0.362$\pm$0.029 \ \\
     NGC~4660  &  40.16  &  40.17  &  39.87  &    9.80 &$<39.45$ & \nodata &\nodata \\
     NGC~4696   &  41.07  &  41.44  &  41.30  &  10.99  &  43.23  &\nodata  &\nodata\\
     NGC~4697   &  40.58  &  41.33  &  40.98  &  10.61  &  40.18  &    8.3$\pm$1.4 \ &  0.148$\pm$0.043 \ \\
     NGC~4915   &  40.92  &  40.95  &  40.76  &  10.34  &$<40.93$ & \nodata &\nodata \\
     NGC~5018   &  41.32  &  42.21  &  41.92  &  10.63  &$<40.59$ & \nodata &\nodata \\
     NGC~5044  &  40.90  &  41.52  &  41.08  &  10.76  &  42.80  &\nodata  &\nodata \\
     NGC~5077  &  41.05  &  41.27  &  40.85  &  10.32  &  40.54  &\nodata  &\nodata \\
     NGC~5322  &  41.08  &  41.75  &  41.42  &  10.73  &  40.27  &\nodata  &\nodata \\
     NGC~5557   &  41.02  &  40.92  &$<40.68$ &  10.90  & \nodata & \nodata &\nodata \\
     NGC~5813  &  40.67  &  40.89  &  40.20  &  10.70  &\nodata  &   16.6$\pm$2.2 \ &  0.059$\pm$0.042 \ \\
     NGC~5846  &  40.88  &  40.93  &  40.53  &  10.72  &  41.71  &   14.2$\pm$2.2 \ &  0.226$\pm$0.051 \ \\
     NGC~5982   &  40.86  &  40.92  &  40.66  &  10.59  &  41.22  & \nodata & \nodata \\
     NGC~6703  &  40.83  &  40.77  &  39.93  &  10.43  &$<40.09$ &    4.8$\pm$0.8 \ &  0.354$\pm$0.034 \ \\
     NGC~7619  &  40.78  &$<39.99$ &$<39.96$ &  10.64  &  41.69  &   15.4$\pm$1.4 \ &  0.291$\pm$0.031 \ \\
     IC~3370  &  41.48  &  42.28  &  41.94  &  10.82  &\nodata  & \nodata &\nodata \\
\enddata

\tablenotetext{a}{Luminosities are calculated with $H_0=70$ km s$^{-1}$ Mpc$^{-1}$.}
\tablenotetext{b}{Ages and metallicities are taken from Thomas et al. 2005, unless otherwise stated.}
\tablenotetext{c}{Age and metallicity for this galaxy are taken from Howell 2005.}
\tablenotetext{d}{Age and metallicity for this galaxy are taken from Sanchez-Blazquez et al. 2006.}
\end{deluxetable}

\clearpage
\begin{figure}%1a
\centering
%\vskip2.in
%%\includegraphics[bb=90 216 522 569,scale=0.9,angle= 270]
%\includegraphics[bb=90 166 522 519,scale=1.0,angle= 0]
\includegraphics{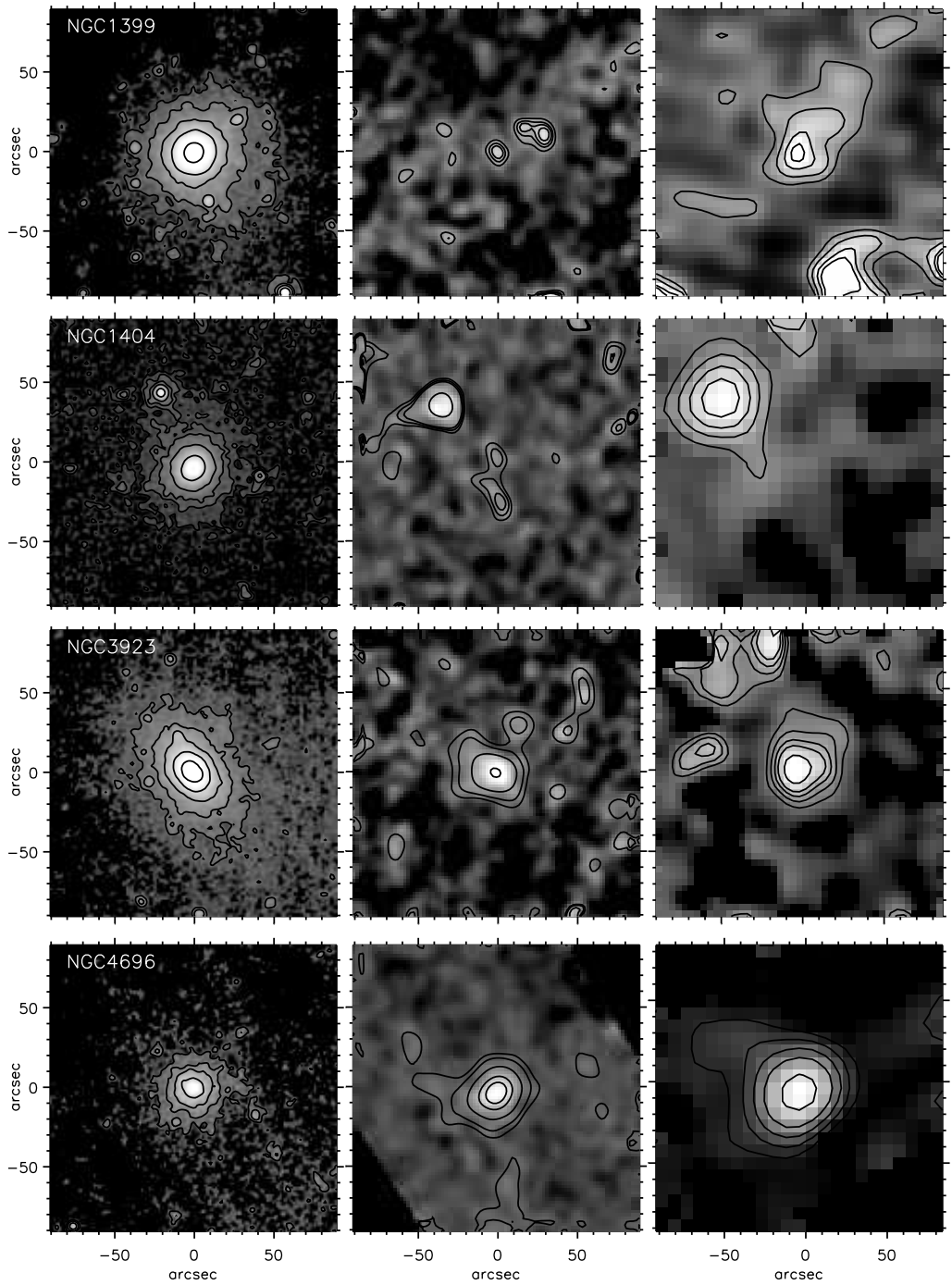}
%\vskip.7in
\caption{
Grey scale images and contour plots of a representative sub-sample of
Ellipticals. Each galaxy is shown in the three MIPS passbands
centered at 24 $\mu$m (left column), 70 $\mu$m (center column),
and 160 $\mu$m (right column). Galaxies are shown in ascending order
from lowest flux density to highest flux density, including
a non detection and sources with extended far-infrared emission.
The contour levels decrease by factors of 2 from the central peak
value.
}
\label{f1_1}
\end{figure}

\setcounter{figure}{0}

\clearpage
\begin{figure}%1b
\centering
%\vskip2.in
\includegraphics{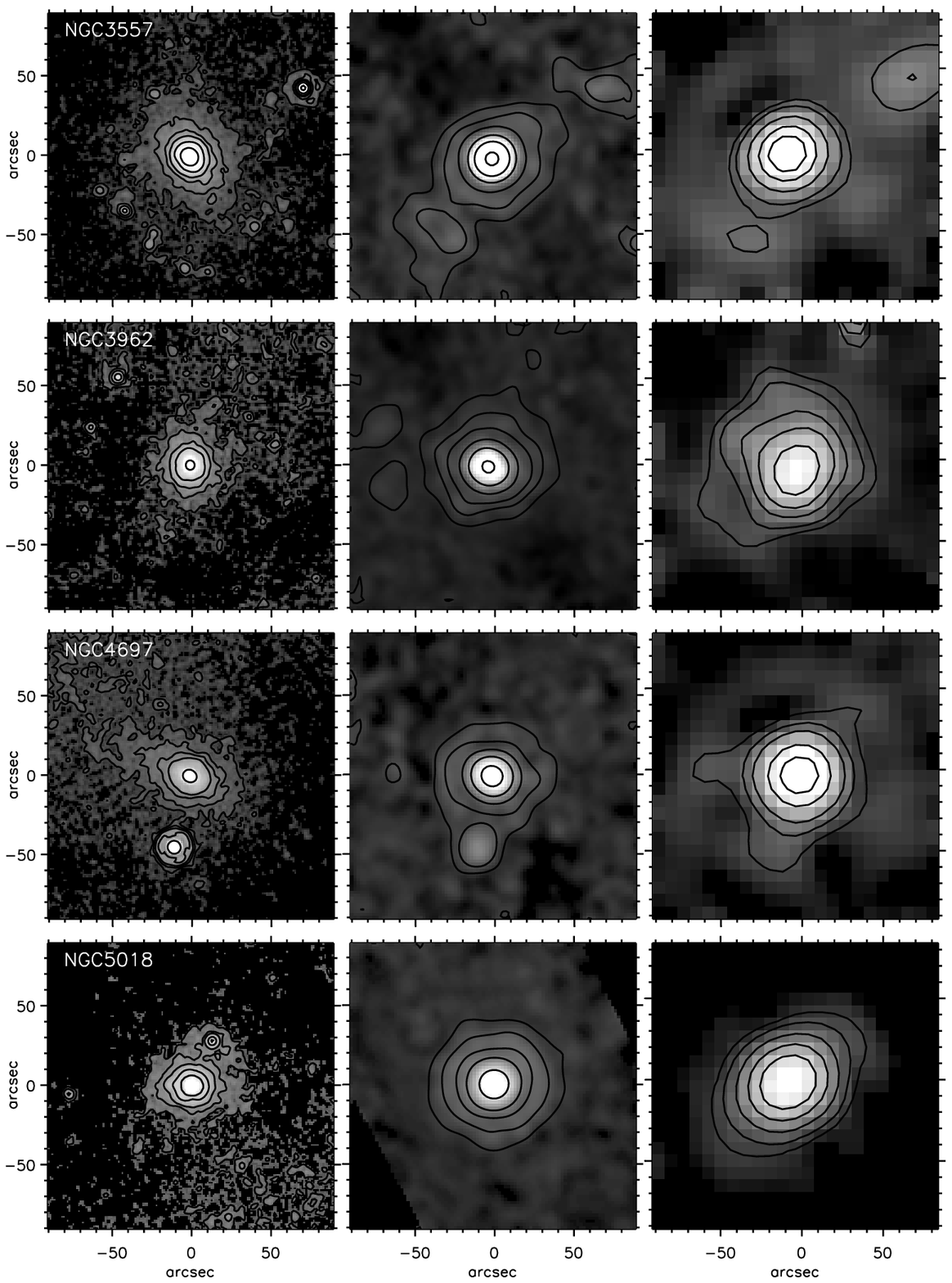}
%\vskip.7in
\caption{
{\it - Continued} }
\label{f1_2}
\end{figure}

\clearpage
\begin{figure}%2
\centering
\vskip2.in
\includegraphics[bb=90 166 522 519,scale=0.8,angle= 0]
{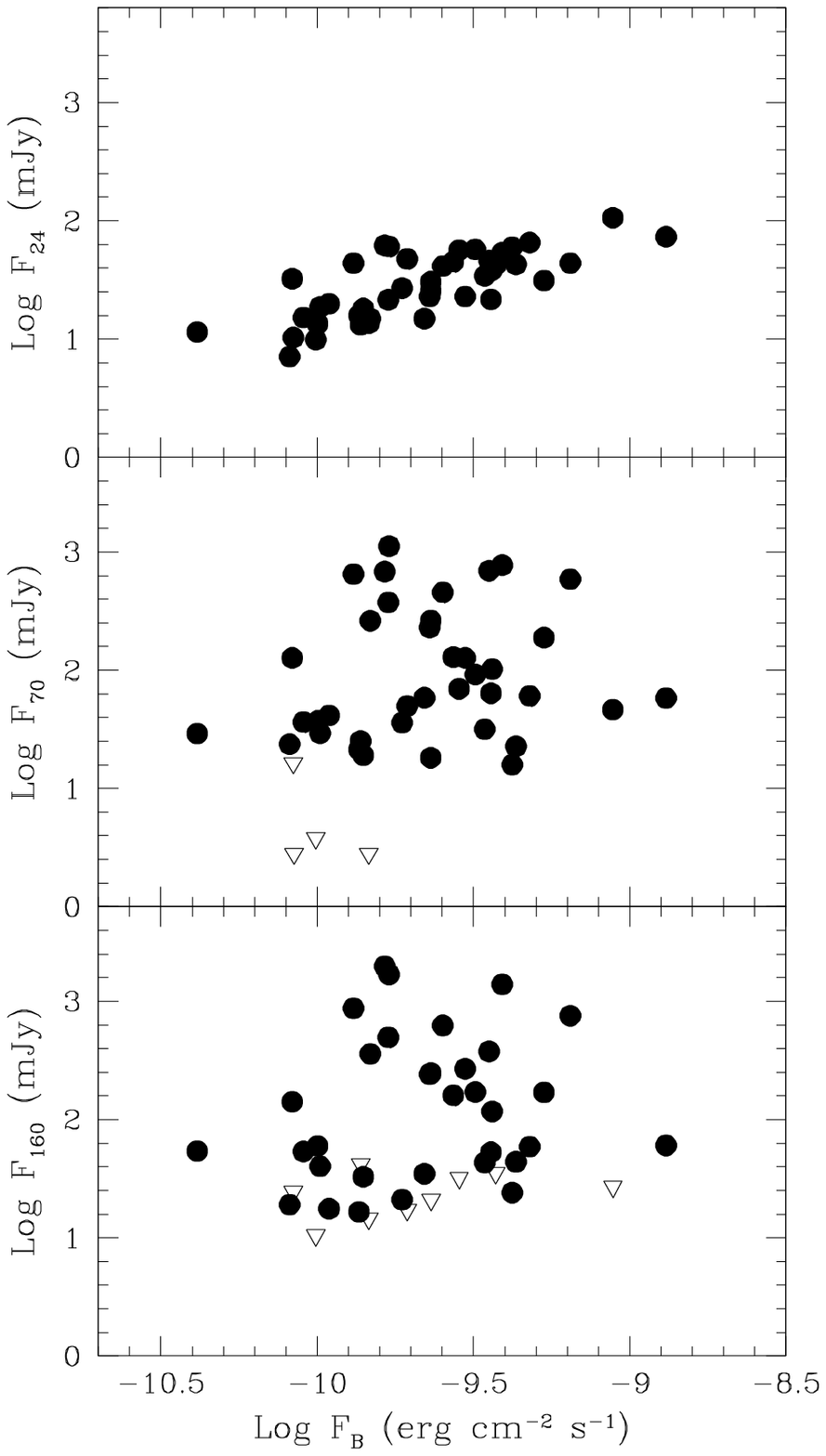}
\vskip.7in
\caption{
Comparison of fluxes at 24, 70 and 160$\mu$m with optical B-band 
fluxes.
Downward pointing open arrows indicate upper limits.
}
\label{f2}
\end{figure}

\clearpage
\begin{figure}%3
\centering
\vskip2.in
\includegraphics[bb=90 166 522 519,scale=0.8,angle= 0]
{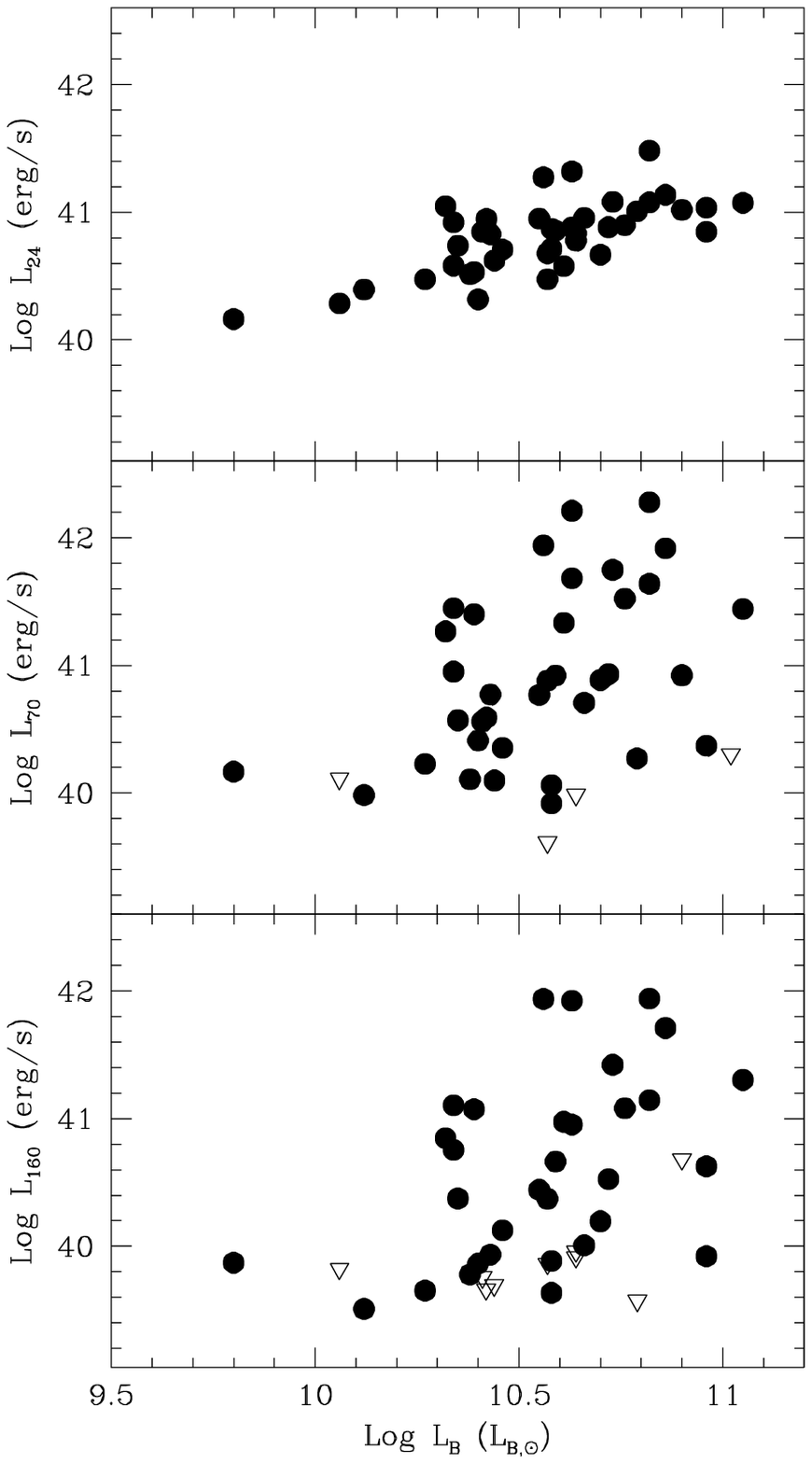}
\vskip.7in
\caption{
Comparison of luminosities at 24, 70 and 160$\mu$m with optical B-band
luminosities.
Downward pointing open arrows indicate upper limits.
}
\label{f3}
\end{figure}

\clearpage
\begin{figure}%4
\centering
\vskip2.in
\includegraphics[bb=90 166 522 519,scale=0.8,angle= 0]
{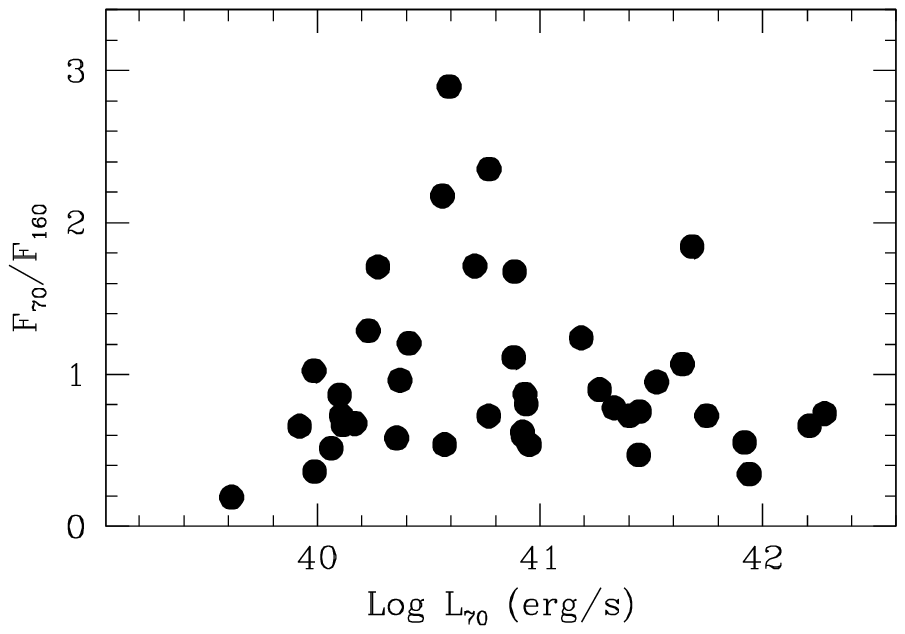}
\vskip.7in
\caption{
Comparision of $F_{70}/F_{160}$, a measure of dust temperature, 
with $L_{70}$.
}
\label{f4}
\end{figure}

\clearpage
\begin{figure}%5
\centering
\vskip2.in
\includegraphics[bb=90 166 522 519,scale=0.8,angle= 0]
{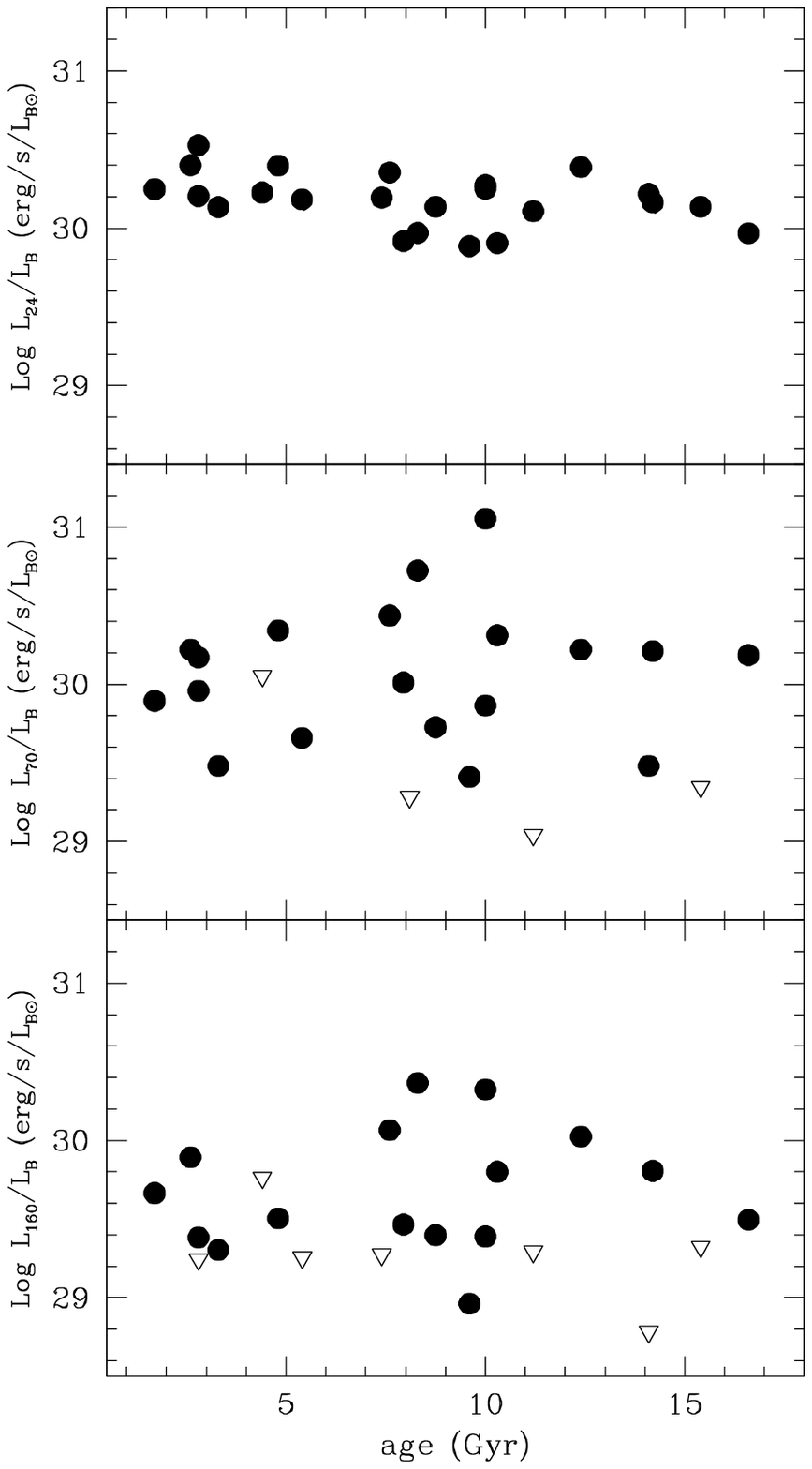}
\vskip.7in
\caption{
FIR to optical luminosity (or flux) ratios plotted against 
apparent stellar age from optical line indices.
}
\label{f5}
\end{figure}

\clearpage
\begin{figure}%6
\centering
\vskip2.in
\includegraphics[bb=90 166 522 519,scale=0.8,angle= 0]
{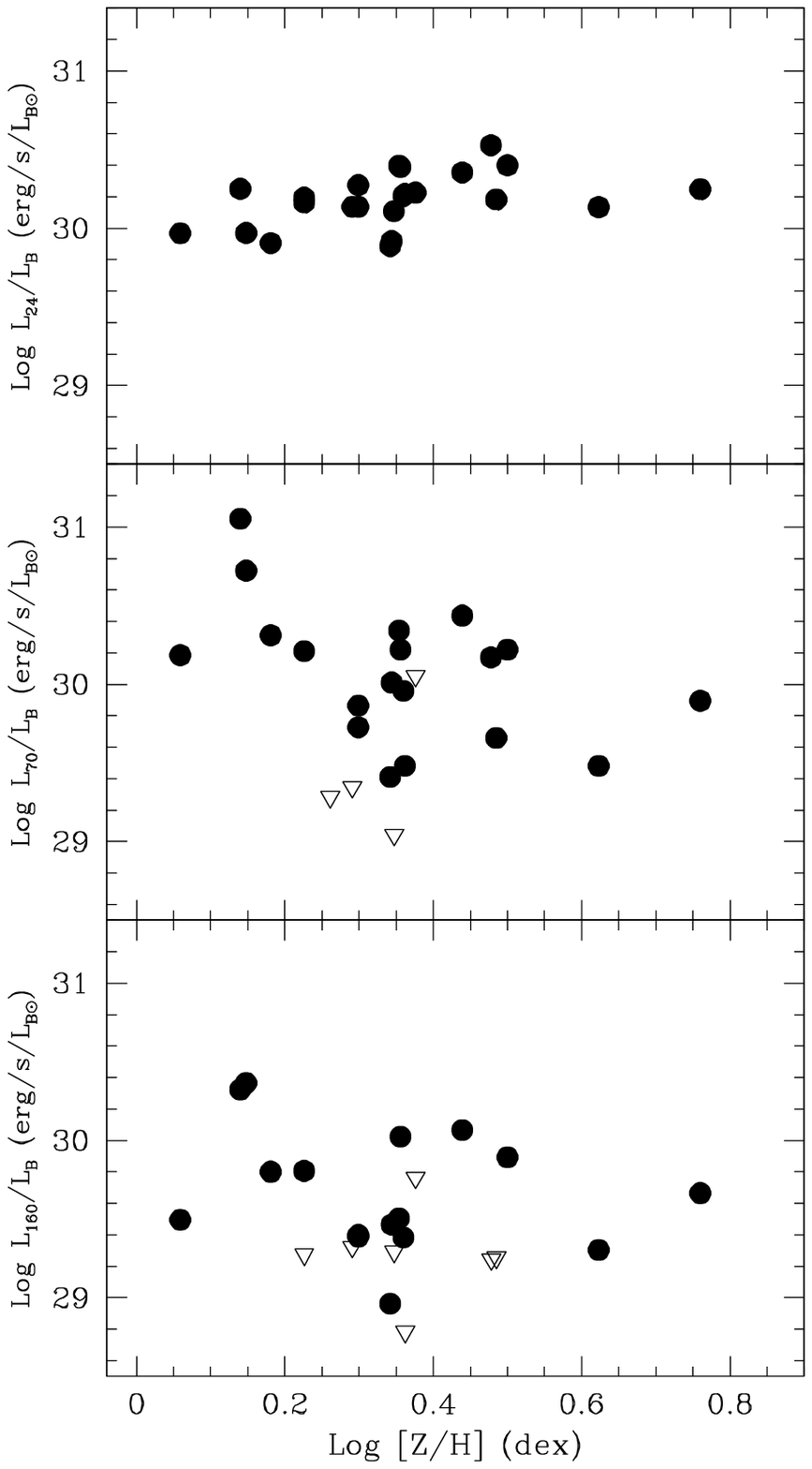}
\vskip.7in
\caption{
FIR to optical luminosity (or flux) ratios plotted against
apparent stellar metallicity from optical line indices.
}
\label{f6}
\end{figure}

\clearpage
\begin{figure}%7
\centering
\vskip2.in
\includegraphics[bb=90 166 522 519,scale=0.8,angle= 0]
{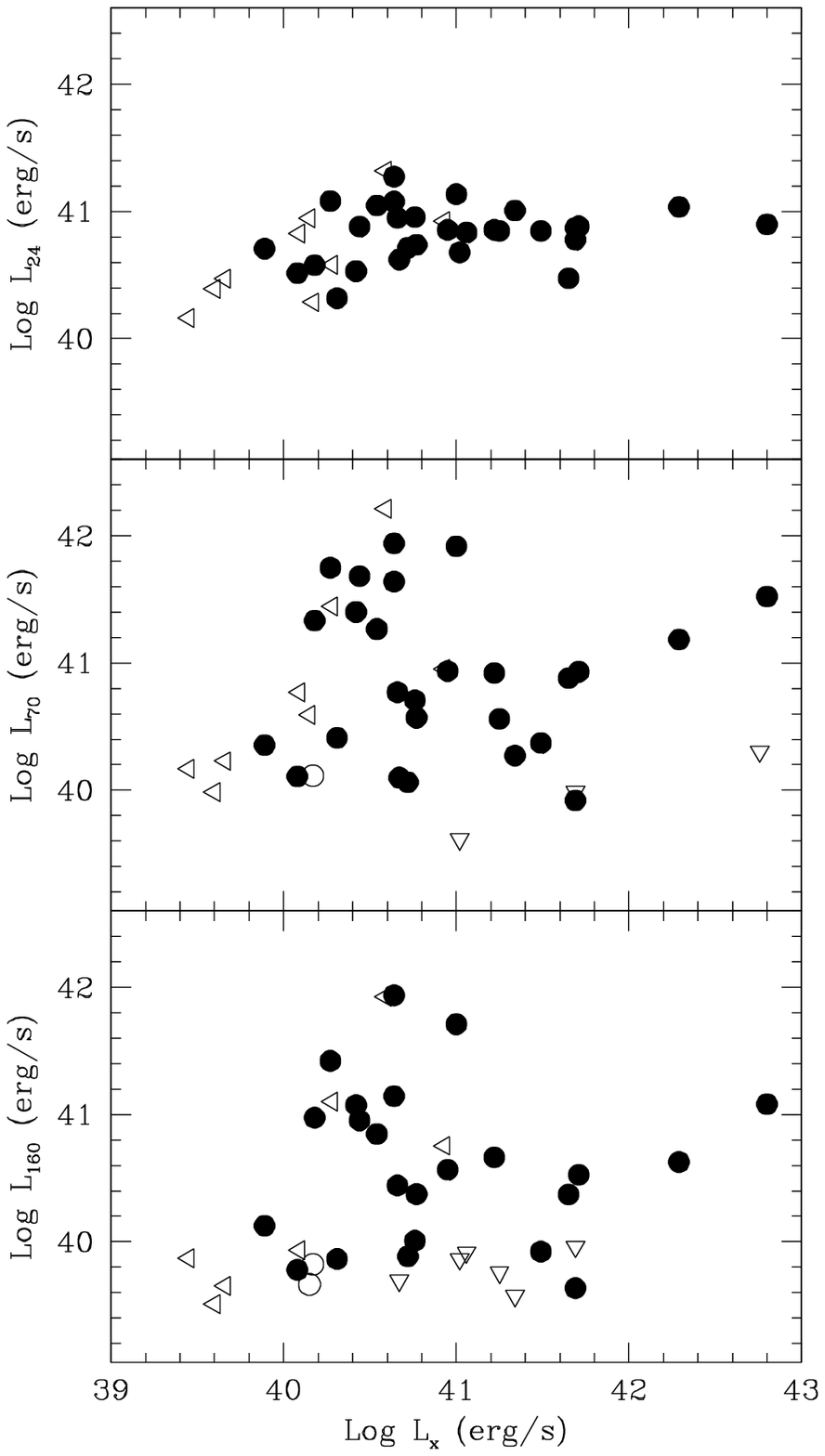}
\vskip.7in
\caption{
Comparison of FIR and X-ray bolometric luminosities. 
}
\label{f7}
\end{figure}

\clearpage
\begin{figure}%8
\centering
\vskip2.in
\includegraphics[bb=90 166 522 519,scale=0.8,angle= 0]
{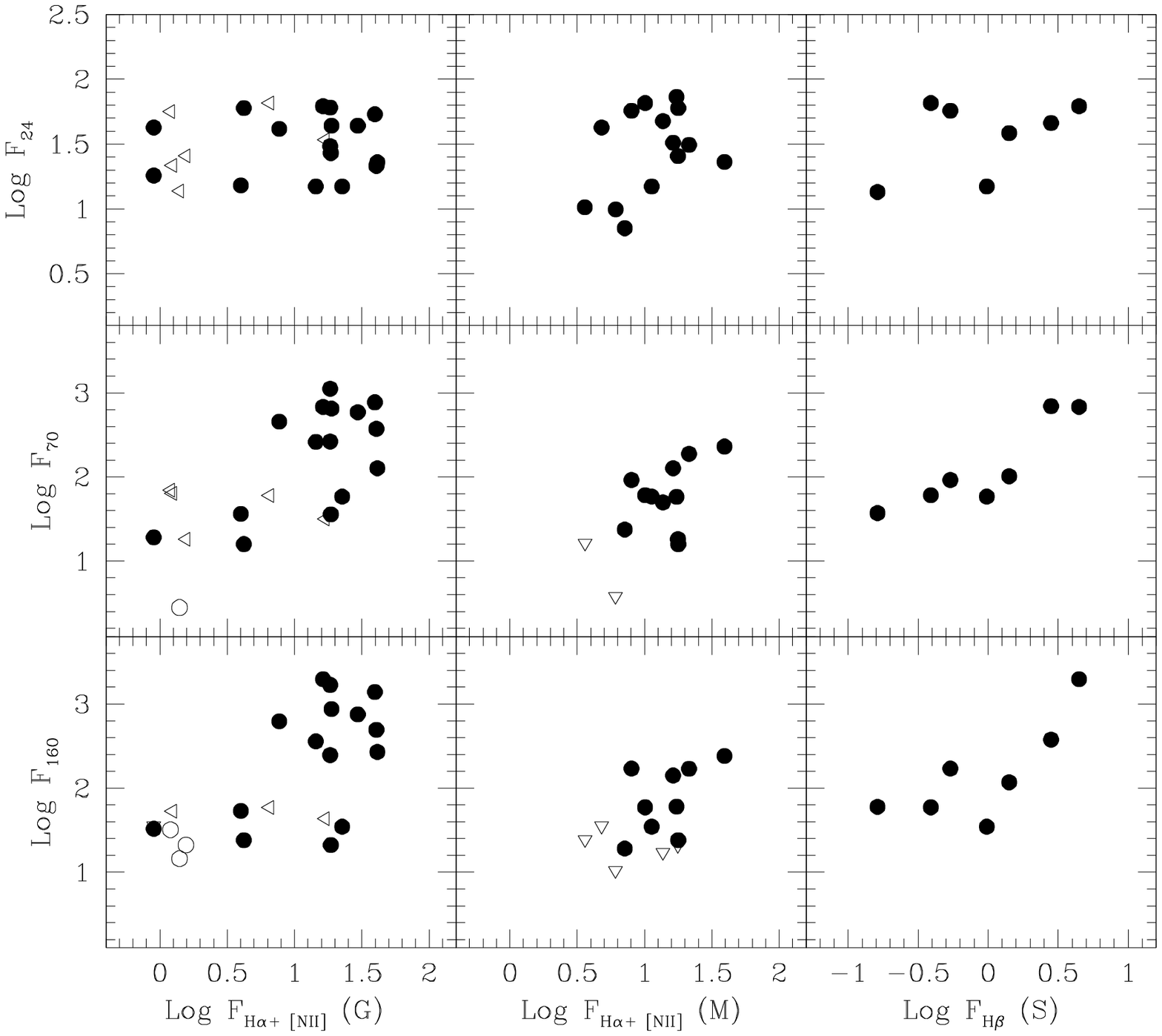}
\vskip.7in
\caption{
FIR fluxes plotted against fluxes in optical line emission. 
>From left to right the three-panel columns refer 
respectively to data from 
Goudfrooij (1994a,b), Macchetto et al. (1996), 
and Sarzi et al. (2006). Open triangles indicate upper limits 
in the direction they are oriented and open circles are 
upper limits in both coordinates.
}
\label{f8}
\end{figure}

\clearpage
\begin{figure}%9
\centering
\vskip2.in
\includegraphics[bb=90 166 522 519,scale=0.8,angle= 270]
{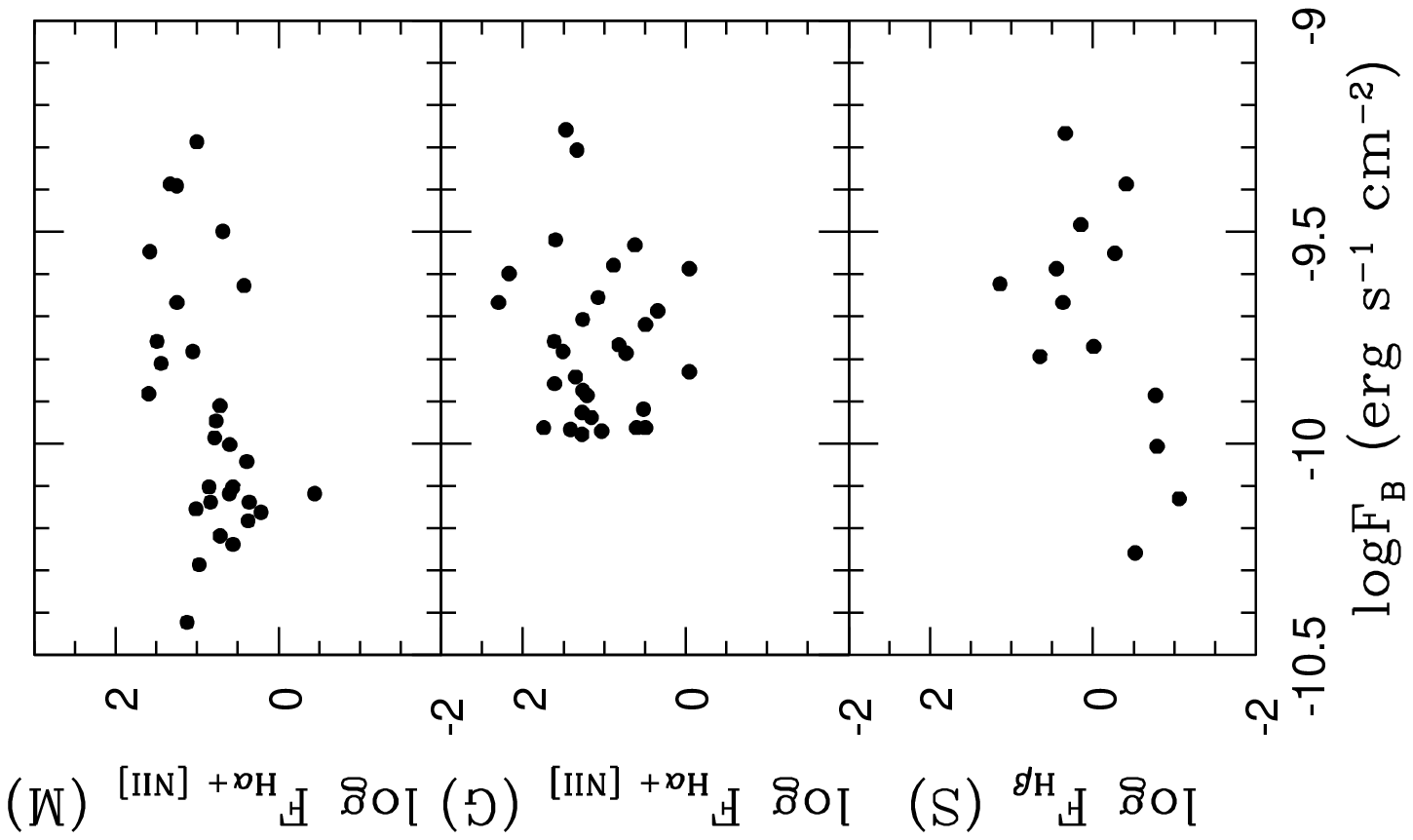}
\vskip.7in
\caption{
Comparison of B-band fluxes with optical emission line fluxes 
in units of $10^{-14}$ erg s$^{-1}$ cm$^{-2}$.
The optical fluxes are taken from 
(G) Goudfrooij (1994a,b), (M) Macchetto et al. (1996),
and (S) Sarzi et al. (2006)
Only non-peculiar E and E/S0 galaxies are plotted. 
}
\label{f9}
\end{figure}

\clearpage
\begin{figure}%10
\centering
\vskip2.in
%\includegraphics[bb=90 216 522 569,scale=0.9,angle= 270]
%1%\includegraphics[bb=90 166 522 519,scale=0.8,angle= 90]
\includegraphics[bb=90 216 522 569,scale=0.8,angle= 90]
{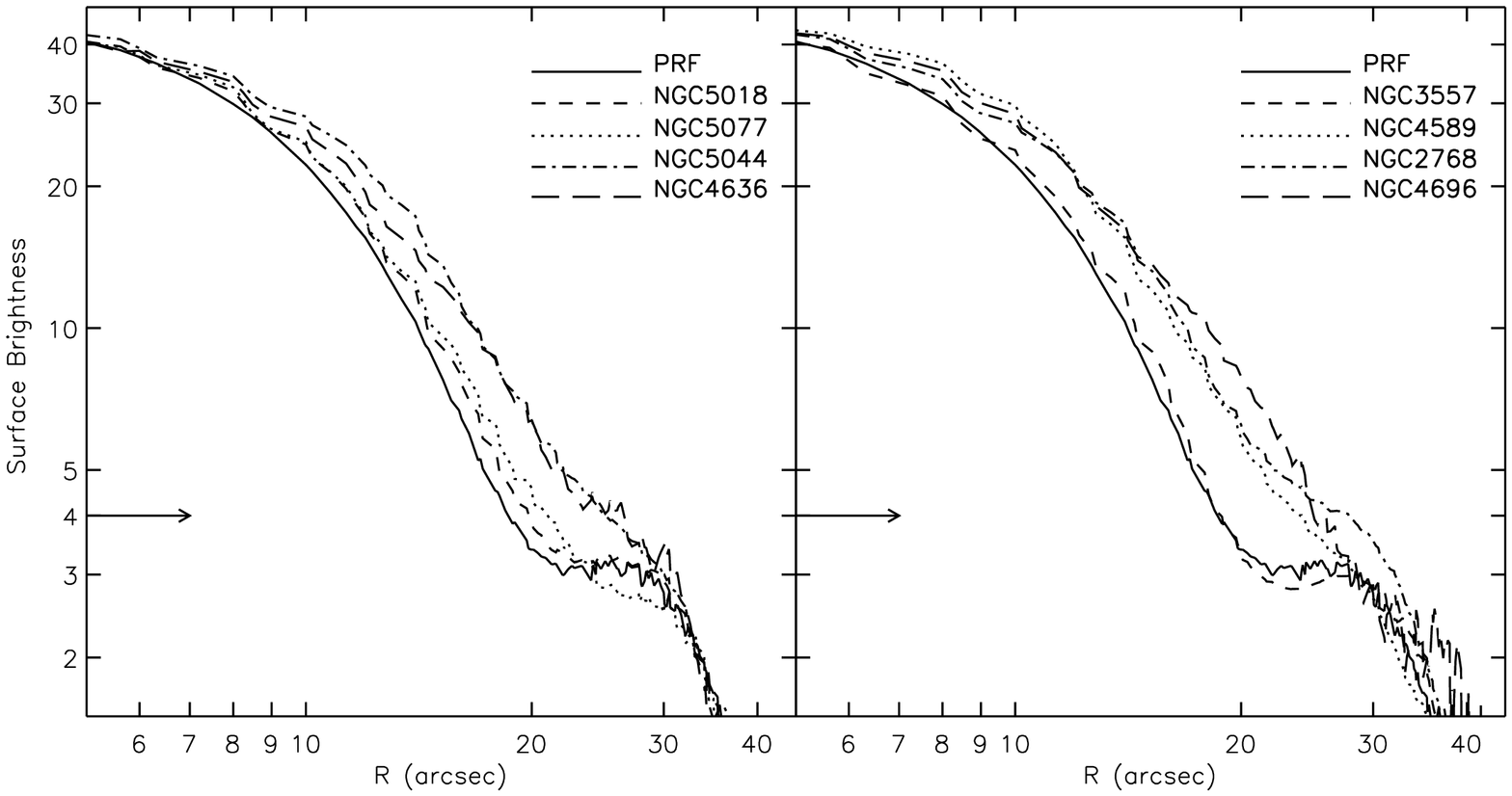}
\vskip.7in
\caption{
Surface brightness distriubtions at 70$\mu$m 
(in arbitrary units) for eight galaxies 
compared to the point response function (PRF) of the MIPS 
detector. 
All curves are normalized to 50. at the center and the 
arrows show the level at 0.1 of the maximum.
}
\label{f10}
\end{figure}

\clearpage
\begin{figure}%11
\centering
\vskip2.in
\includegraphics[bb=90 166 522 519,scale=0.8,angle= 0]
{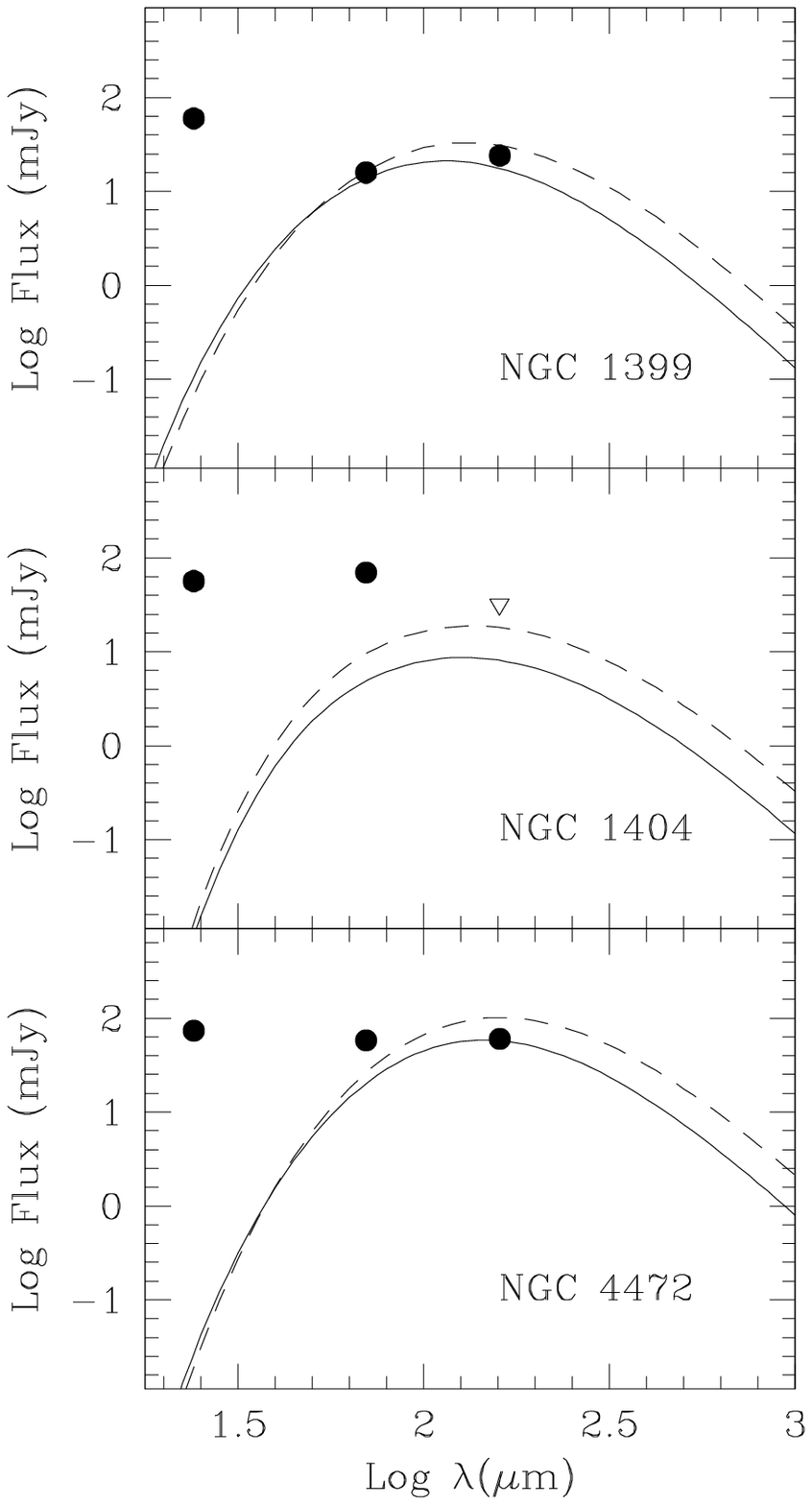}
\vskip.7in
\caption{
Comparison of 24, 70 and 160$\mu$m fluxes with the our 
model spectral energy distributions. 
The open triangle is an upper limit of $F_{160}$ for NGC 1404. 
The solid and dashed lines
correspond respectively to emission from
grains with original maximum sizes $a_{max} = 0.3\mu$m and 1.0$\mu$m.
}
\label{f11}
\end{figure}

\clearpage
\appendix
\section{Nature of FIR-luminous Galaxies}

In this appendix we briefly review other
observations of galaxies in Figure 3 and Table 2 having 
the most luminous FIR emission.
Our objective is to determine if elliptical galaxies with
the most luminous FIR emission can be regarded as unusual
or abnormal and, if so, what is the maximum FIR luminosity
of sample galaxies that can be regarded as normal.
In the following we emphasize observations that
are consistent with large FIR emission -- evidence of
large ($\sim 10$ kpc) extended dusty disks, 
detections of large masses of cold gas or recent star formation, 
designations
as an S0 galaxy which often contain cold, rotationally supported
gaseous disks, strong radio emission, galaxies with nearby 
dusty galaxies or
stars that may have contaminated some observations,
evidence of a recent merger, etc. --  but
we do not
mention attributes that are found in many typical ellipticals:
extended optical emission lines perhaps with a LINER spectrum,
small central ($\lta 1$ kpc) dust clouds that obscure starlight,
weak central radio sources, X-ray emission $L_x$ which is
highly variable among ellipticals, evidence of old mergers, etc.
Galaxies are listed in descending order of $L_{70}$.

\noindent\vskip.15in
{\bf IC 3370:} Classified as an elliptical in RC2 an RSA, but regarded
by Jarvis (1987) and Samurovic \& Danziger (2005)
as an S0 or S0pec with an extremely boxy profile.
A bright IRAS source in FIR.
Richter et al. (1994) detect an HI mass of $5 \times 10^8$ $M_{\odot}$.
Contains a prominent dust lane in the inner regions
with associated H$\alpha$ + [NII]
emission and possible evidence of ongoing star formation
(Michard 2006).
Leonardi \& Worthey (2000) and Tantalo \& Chiosi (2004)
suggest an intermediate age population
for the central regions.

\noindent\vskip.15in
{\bf NGC 5018:} Forms a pair with NGC 5022 7.2$^{\prime}$ away.
Carollo \& Danziger (1994) describe a very strong dust lane that
obscures the innermost regions that may contain recent star
formation.
They also suggest that the galaxy is very
flat (about E6). 
Detected by IRAS at 60 and 100$\mu$m.
Xilouris et al. (2004) find excess MIR emission at 6.7 and 15$\mu$m,
perhaps suggesting heating by recent star formation.

\noindent\vskip.15in
{\bf NGC 2974:} A rapidly rotating E4 galaxy
(Emsellem et al. 2004).
Strong IRAS emission.
Kim et al. (1988) observes a massive HI disk, $M_{HI} = 2 \times 10^9$
$M_{\odot}$ (with $D = 38$ Mpc).
Sarzi et al. (2006) observe a nuclear bar with two spiral arms
emerging.
A bright star is seen in NED images 0.7$^{\prime}$ to SW.

\noindent\vskip.15in
{\bf NGC 4125:} Filho et al. (2002) regard this as an E6pec galaxy.
Wickland et al. (1995) detect tentative molecular hydrogen emission
with mass $M_{H2} = 7 \times 10^7$ $M_{\odot}$.
Verdoes Kleijn \& de Zeeuw (2005) observe a central dust lane
of size 160 pc (D = 26 Mpc).

\noindent\vskip.15in
{\bf NGC 5322:} Carollo et al. (1997) describe a completely obscuring
dust lane (perpendicular to the radio jet) that hides the center
of the galaxy.
IRAS detection at 60 and 100$\mu$m.
Proctor \& Sansom (2002) and Denicolo et al. (2005)
find evidence of a $\sim 2-4$ Gyr old young stellar
population.

\noindent\vskip.15in
{\bf NGC 2768:} Referred to as an E6* in RC3,
this is almost certainly an S0 galaxy (Funes et al. 2002)
with faint dust patches surrounding the galactic bulge.
Significant dust extinction is visible
along the minor axis (Michard 1999).
Bertola et al. (1992) notes that the ionized gas is dynamically
decoupled from the stars, which may be related to
gas orbiting in a polar ring (Mollenhoff et al. 1992;
Fried \& Illingworth 1994).
Wiklind et al. (1995) observed CO emission with an estimated
molecular mass of $\sim 2\times 10^7$ $M_{\odot}$.

\noindent\vskip.15in
{\bf NGC 3557:} An E3 galaxy containing an FRI radio source
having a double-tailed radio source with a jet and
central knot (Birkinshae \& Davies 1985).
Detected by IRAS at 60 and 100$\mu$m.
Rampazzo et al. (2005) report a possible dust ring near the center
and Goudfrooij (1994b) finds an unusual optical emission line
image associated with the minor axis.

\noindent\vskip.15in
{\bf NGC 5044:} Detected by IRAS at 60 and 100$\mu$m.
Very bright and asymmetrically extended optical emission
(Ferrari, et al. 1999; Goudfrooij et al. 1994a).

\noindent\vskip.15in
{\bf NGC 3962:} The H$\alpha$+[NII] image shows two sub-systems: 
an elongated (dusty)
central component strongly misaligned with the stellar
isophotes extending to $\sim 7$ kpc with regular kinematics
and a more extended arm-like structure associated with bluer
stars
(Buson et al. 1993; Goudfrooij, et al. 1994b;
Zeilinger et al. 1996).
Detected by IRAS at 60 and 100$\mu$m.

\noindent\vskip.15in
{\bf NGC 4589:} Prominent dust lane near minor axis extending
to 20$^{\prime\prime}$ (Goudfrooij, et al. 1994a,b)
Complex gas and stellar kinematics
suggest a merger remnant (Mollenhoff \& Bender 1989).
Detected by IRAS at 60 and 100$\mu$m.

\noindent\vskip.15in
{\bf NGC 4697:} An E6 galaxy, but may be an S0
(Koprolin \& Zeilinger 2000). Sofue and Wakamatsu (2003) detect
a molecular mass of $3 \times 10^7$ $M_{\odot}$.

\noindent\vskip.15in
{\bf NGC 5077:} The E3 designation in RC2 is wrong, this is
an S0 galaxy (Sandage \& Bedke 1994).

\clearpage
%\appendix
\section{Estimated FIR Emission from Central Optically Thick Cloud}

In this Appendix we determine if a large fraction
of the FIR emission from elliptical galaxies can come from small,
dense, dusty clouds near the centers of the galaxy, or,
as we have generally assumed, comes from FIR-emitting
dust diffusely distributed throughout the inner galaxy.
Specifically, can reprocessed starlight from internal and external
stars heat small dusty regions sufficiently to account for the
large FIR luminosities shown in Figure 3?
We describe a simple calculation to estimate the maximum
luminosity at 70$\mu$m $L_{70}$ expected from a small
optically thick dust cloud, including dust heating by both
internal and external starlight.

Consider a spherical cloud of dusty cold gas with radius $r_c$
placed at the center of an elliptical galaxy having a
de Vaucouleurs luminosity profile.
Since the stellar density is greatest at the center,
placing the cloud there maximizes the
energy from internal starlight that can be absorbed and reemitted
at infrared wavelengths.
The mean intensity of starlight in an (optically thin) elliptical
galaxy also is greatest at the galactic center,
where the stellar flux incident on the
dusty cloud from the outside is also maximal.

Without investigating in detail how the dust is heated and
reemitted in the infrared from small, dense clouds, 
we assume that this type of reprocessing can occur.
To estimate the fraction of the the FIR that appears in the
{\it Spitzer} 70$\mu$m pass band, we assume that the overall
infrared SED can be approximated with a modified blackbody,
with emissivity $j_{nu} \propto \nu^{\beta}B_{nu}(T_d)$
where $\beta$ depends on the properties of the dust
and $T_d$ is the dust temperature.
We assume $\beta = 1$ and $T_d = \langle T_d \rangle = 27.9$K,
the mean values in our SED fits to
ISO ({\it Infrared Space Observatory}) FIR data
for elliptical galaxies (Temi et al. 2004).
The fraction of FIR emission that emerges in the
{\it Spitzer} 70$\mu$m passband 
(width $\Delta \lambda_{70} = 19\mu$m) is
\begin{equation}
{L_{70} \over L_{FIR}} =
{ \int_{\Delta \nu_{70}}
\nu^{\beta}B_{\nu}(\langle T_d \rangle) d \nu
\over \int_{0}^{\infty} \nu^{\beta}B_{\nu}(\langle T_d \rangle) 
d \nu }
= 0.147
\end{equation}
where $\Delta \nu_{70} = c \Delta \lambda_{70}/\lambda_{70}^2$.

To be specific we consider a typical elliptical in Figure 3
with $\log L_{B} = 10.6$.
Based on the galaxy-size observations of Shen et al. (2003),
we assume a typical effective radius for this
galaxy, $r_e = 7.67$ kpc.
To determine the total B-band luminosity of stars
inside the dusty cloud,
we integrate the de Vaucouleurs profile from the center
to radius $r_c$.
To convert the B-band luminosity to bolometric luminosity
of an old single stellar population (SSP),
we assume solar abundance for the stars, a Kroupa IMF and
an age of $10^{10}$ yrs, resulting in
$(L_{B}/L_{B\odot})/(L_{bol}/L_{bol,\odot}) = 0.396$
(private communication from Justin Harker).
With this information we can estimate the total
stellar luminosity produced
within a dusty cloud of any chosen radius $r_c$
and the fraction of this energy that is observed in the
{\it Spitzer} 70$\mu$m passband, $L_{70}(int)$.

The dusty cloud can also be heated 
externally by stellar radiation incident
on its outer surface.
To estimate the flux of stellar radiation incident on
the dusty cloud we first determine
the mean B-band mean intensity within 10-100 parsecs of the
galactic center,
$J_B \approx 0.03$ erg cm$^{-2}$ s$^{-1}$ ster$^{-1}$,
found by integrating over
the de Vaucouleurs emissivity profile corresponding to $L_B = 10^{10.6}$
$L_{B\odot}$ and $r_e = 7.67$ kpc.
The B-band flux on the surface of the cloud is then
$F_B = \pi J_B$.
The bolometric flux is larger by a factor
$F_{bol,\odot}/F_{B,\odot} = 1/0.396$
so the rate that radiative energy from external galactic
stars is absorbed by the dusty cloud is
$L_{70}(ext) = a4 \pi r_c^2 (F_{B}/0.396)$
where $a = 0.5$ is the assumed albedo.

Figure B1 shows how the 70$\mu$m luminosities $L_{70}$(int)
and $L_{70}$(ext) vary with the radius $r_c$ of the dusty
cloud where we see that
$L_{70}(ext) < L_{70}(int)$.
But the important result is that clouds of size
$r_c \sim 100$ pc are required to produce luminosities
$L_{70} \sim 10^{41}$ erg s$^{-1}$ typical of those
plotted in the central panel of Figure 3.
In many of the elliptical galaxies in Figure 3, a
$\sim 100$ parsec dark cloud would be easily visible even
without HST resolution.
We conclude that it is very unlikely that the large
excess FIR emission in Figure 3 above that 
predicted by our simple
model for recently produced diffuse dust, can be explained
by emission from small dense clouds even under the most
favorable circumstances.

More importantly, the approximation of spherical clouds is
seriously in error.
The gas temperatures associated
with these dusty clouds ($\sim 100$K?) are
much less than the virial temperature of the stars,
$T_* \sim 10^6$ K required to support a spherical
cloud as we hypothesized here.
The gas temperature within the dusty cloud is insufficient to
provide pressure support against the stellar potential
due to embedded stars or the external hot gas pressure. 
Consequently, we expect that most of the nuclear dust
clouds observed in ellipticals are highly flattened,
nearly two-dimensional pancakes or disks.
In this case we imagine that the $L_{70}$(ext) luminosities
in Figure B1 are more relevant, further reducing the predicted
$70\mu$m emission.
We conclude that most of the FIR emission observed at
70$\mu$m and 160$\mu$ is diffusely distributed, not
concentrated in dense clouds similar to those seen
in absorption by many optical observers.

\clearpage
\begin{figure}%12
\centering
\vskip2.in
\includegraphics[bb=90 166 522 519,scale=0.7,angle= 270]
{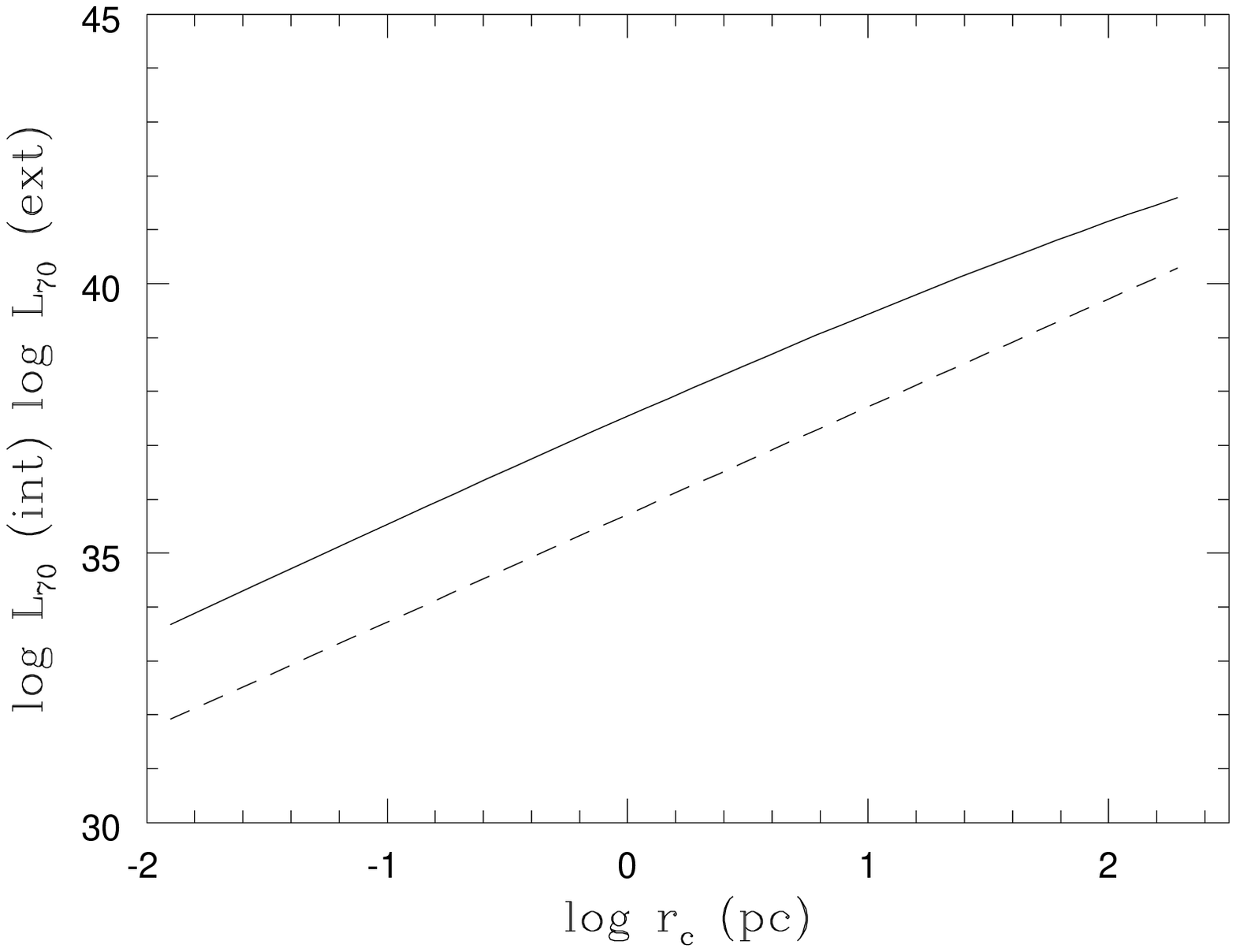}
\vskip.7in
\caption{
Estimated 70$\mu$m luminosity emitted from an optically thick 
dust sphere located a the center of a elliptical galaxy 
having a de Vaucouleurs stellar luminosity with $L_B = 10^{10.6}$
$L_{B\odot}$ and $r_e = 7.67$ kpc.
As the cloud radius $r_{c}$ is varied, 
the two curves show the 70$\mu$m luminosity from reprocessed starlight 
emitted (and absorbed) 
within the dust cloud $L_{70}(int)$ ({\it solid line}) and 
the reprocessed luminosity absorbed by starlight incident on the sphere
from galactic stars beyond $r_{c}$, $L_{70}(ext)$ 
({\it dashed line}).
}
\label{f12}
\end{figure}

\end{document}